\documentclass[aps,preprint]{revtex4}

\usepackage{graphicx}

\newcommand{\bea}{\begin{eqnarray}}
\newcommand{\eea}{\end{eqnarray}}
\newcommand{\be}{\begin{equation}}
\newcommand{\ee}{\end{equation}}
\newcommand{\Dslash}{D\!\!\!\!/\,}

\newcommand{\slashed}[1]{#1\!\!\!/}

\newcommand{\lb}{\lambda}

\newcommand{\gpm}{\gamma_{\pm}}
\newcommand{\gmp}{\gamma_{\mp}}
\newcommand{\gf}{\gamma_{5}}

\newcommand{\ivec}[1]{\stackrel{\leftarrow}{#1}} 

\begin{document}


\title{Instantons and the Spin of the Nucleon}

\author{T.~Sch\"afer$^{1,3}$ and V.~Zetocha$^{1,2}$}

\affiliation{
$^1$Department of Physics, North Carolina State University,
Raleigh, NC 27696\\
$^2$Department of Physics, SUNY Stony Brook, Stony Brook, NY 11794\\
$^3$Riken-BNL Research Center, Brookhaven National Laboratory,
Upton, NY 11973}

\begin{abstract}
  
 Motivated by measurements of the flavor singlet axial 
coupling constant of the nucleon in polarized deep inelastic 
scattering we study the contribution of instantons to OZI 
violation in the axial-vector channel. We consider, in particular, 
the $f_1-a_1$ meson splitting, the flavor singlet and triplet
axial coupling of a constituent quark, and the axial coupling
constant of the nucleon. We show that instantons provide a short 
distance contribution to OZI violating correlation functions
which is repulsive in the $f_1$ meson channel and adds 
to the flavor singlet three-point function of a constituent 
quark. We also show that the sign of this contribution
is determined by positivity arguments. We compute long 
distance contributions using numerical simulations of the 
instanton liquid. We find that the iso-vector axial coupling 
constant of a constituent quark is $(g_A^3)_Q=0.9$ and that 
of a nucleon is $g_A^3=1.28$, in good agreement with experiment. 
The flavor singlet coupling of a quark is close to one, while 
that of a nucleon is suppressed, $g_A^0=0.77$. However, this 
number is larger than the experimental value $g_A^0=(0.28-0.41)$.

\end{abstract}

\maketitle

\newpage

\section{Introduction}
\label{sec_intro}

 The current interest in the spin structure of the nucleon
dates from the 1987 discovery by the European Muon Collaboration
that only about 30\% of the spin of the proton is carried by 
the spin of the quarks \cite{Ashman:1987hv}. This results is
surprising from the point of view of the naive quark model, 
and it implies a large amount of OZI (Okubo-Zweig-Iizuka rule)
violation in the flavor singlet axial vector channel. The axial 
vector couplings of the nucleon are related to polarized quark 
densities by
\bea
 g_A^3 &=& \Delta u - \Delta d, \\
 g_A^8 &=& \Delta u + \Delta d - 2\Delta s,\\
 g_A^0 & \equiv & \Delta \Sigma = \Delta u + \Delta d + \Delta s.
\eea
The first linear combination is the well known axial vector coupling 
measured in neutron beta decay, $g_A^3=1.267\pm 0.004$.
The hyperon decay constant is less well determined. A conservative
estimate is $g_A^8= 0.57\pm 0.06$. Polarized deep inelastic
scattering is sensitive to another linear combination of the 
polarized quark densities and provides a measurement of the 
flavor singlet axial coupling constant $g_A^0$. Typical results 
are in the range $g_A^0=(0.28-0.41)$, see \cite{Filippone:2001ux} 
for a recent review. 

 Since $g_A^0$ is related to the nucleon matrix element of the 
flavor singlet axial vector current many authors have speculated 
that the small value of $g_A^0$ is in some way connected to the 
axial anomaly, see \cite{Dorokhov:ym,Bass:1993bs,Bass:2003vp} 
for reviews. The axial anomaly relation
\be
\label{anom}
\partial^\mu A_\mu^0 = \frac{N_fg^2}{16\pi^2}
 G_{\mu\nu}^a \tilde{G}_{\mu\nu}^a
 + \sum_f 2m_f\bar{q}_fi\gamma_5q_f
\ee
implies that matrix elements of the flavor singlet axial 
current $A_\mu^0$ are related to matrix elements of the 
topological charge density. The anomaly also implies that
there is a mechanism for transferring polarization from 
quarks to gluons. In perturbation theory the nature of the 
anomalous contribution to the polarized quark distribution 
depends on the renormalization scheme. The first moment of 
the polarized quark density in the modified minimal
subtraction $(\overline{MS})$ scheme is related to the 
first moment in the Adler-Bardeen $(AB)$ scheme by 
\cite{Altarelli:1988nr}
\be
\Delta \Sigma_{\overline{MS}} = \Delta \Sigma_{AB} 
 -N_f\frac{\alpha_s(Q^2)}{2\pi} \Delta G(Q^2),
\ee
where $\Delta G$ is the polarized gluon density. Several 
authors have suggested that $\Delta \Sigma_{AB}$ is more 
naturally associated with the ``constituent'' quark spin 
contribution to the nucleon spin, and that the smallness 
of $\Delta \Sigma_{\overline{MS}}$ is due to a cancellation 
between $\Delta\Sigma_{AB}$ and $\Delta G$. The disadvantage
of this scheme is that $\Delta\Sigma_{AB}$ is not associated
with a gauge invariant local operator \cite{Jaffe:1989jz}. 

 Non-perturbatively the anomaly implies that $g_A^0=\Delta
\Sigma$ can be extracted from nucleon matrix elements 
of the topological charge density $g^2G^a_{\mu\nu}
\tilde{G}^a_{\mu\nu}/(32\pi^2)$ and the pseudoscalar 
density $m\bar{\psi}i\gamma_5\psi$. The nucleon matrix
element of the topological charge density is not known, 
but the matrix element of the scalar density $g^2G^a_{\mu\nu}
G^a_{\mu\nu}$ is fixed by the trace anomaly. We have 
\cite{Shifman:zn}
\be 
\langle N(p)| \frac{g^2}{32\pi^2}G^a_{\mu\nu}G^a_{\mu\nu}
 | N(p')\rangle = C_S(q^2)m_N\bar{u}(p)u(p'),
\ee
with $C_S(0)=-1/b$ where $b=11-2N_f/3$ is the first coefficient 
of the QCD beta function. Here, $u(p)$ is a free nucleon spinor. 
Anselm suggested that in an instanton model of the QCD vacuum 
the gauge field is approximately self-dual, $G^2=\pm G\tilde{G}$, 
and the nucleon coupling constants of the scalar and pseudoscalar 
gluon density are expected to be equal, $C_S(0)\simeq C_P(0)$
\cite{Anselm:1992wz}, see also \cite{Kuhn:1990df}. Using $g_A^0=
N_f C_P(0)$ in the chiral limit we get $g_A^0 \simeq - N_f/b \simeq 
-0.2$, which is indeed quite small. 

 A different suggestion was made by Narison, Shore, and 
Veneziano \cite{Narison:hv}. Narison et al.~argued that 
the smallness of $\Delta \Sigma=g_A^0$ is not related
to the structure of the nucleon, but a consequence
of the $U(1)_A$ anomaly and the structure of the 
QCD vacuum. Using certain assumptions about the 
nucleon-axial-vector current three-point function 
they derive a relation between the singlet and octet 
matrix elements,
\be
\label{vs}
 g_A^0 = g_A^8 \frac{\sqrt{6}}{f_\pi} \sqrt{\chi'_{top}(0)}.
\ee
Here, $f_\pi=93$ MeV is the pion decay constant and 
$\chi'_{top}(0)$ is the slope of the topological charge 
correlator
\be 
\label{chi}
\chi_{top}(q^2) = \int d^4x\, e^{iqx} \,
 \langle Q_{top}(0) Q_{top}(x) \rangle ,
\ee
with $Q_{top}(x)=g^2G^a_{\mu\nu}\tilde{G}^a_{\mu\nu}/(32\pi^2)$.
In QCD with massless fermions topological charge is screened
and $\chi_{top}(0)=0$. The slope of the topological charge 
correlator is proportional to the screening length. In QCD 
we expect the inverse screening length to be related to the 
$\eta'$ mass. Since the $\eta'$ is heavy, the screening 
length is short and $\chi'_{top}(0)$ is small. Equation 
(\ref{vs}) relates the suppression of the flavor singlet 
axial charge to the large $\eta'$ mass in QCD. 

 Both of these suggestions are very interesting, but the status 
of the underlying assumptions is somewhat unclear. In this 
paper we would like to address the role of the anomaly in 
the nucleon spin problem, and the more general question
of OZI violation in the flavor singlet axial-vector channel, 
by computing the axial charge of the nucleon and the axial-vector 
two-point function in the instanton model. There are several reasons 
why instantons are important in the spin problem. First of all, 
instantons provide an explicit, weak coupling, realization of 
the anomaly relation equ.~(\ref{anom}) and the phenomenon 
of topological charge screening \cite{Schafer:1996wv,Shuryak:1994rr}.
Second, instantons provide a successful phenomenology
of OZI violation in QCD \cite{Schafer:2000hn}. Instantons
explain, in particular, why violations of the OZI rule 
in scalar meson channels are so much bigger than OZI
violation in vector meson channels. And finally, the 
instanton liquid model gives a very successful description
of baryon correlation functions and the mass of the 
nucleon \cite{Schafer:1993ra,Diakonov:2002fq}.

 This paper is organized as follows. In Sect.~\ref{sec_anom}
we review the calculation of the anomalous contribution to the 
axial-vector current in the field of an instanton. In 
Sect.~\ref{sec_pia} and \ref{sec_gaq} we use this result 
in order to study OZI violation in the axial-vector correlation 
function and the axial coupling of a constituent quark. Our 
strategy is to compute the short distance behavior of the 
correlation functions in the single instanton approximation 
and to determine the large distance behavior using numerical 
simulations. In Sect.~\ref{sec_gan} we present numerical 
calculations of the axial couplings of the nucleon and in 
Sect.~\ref{sec_sum} we discuss our conclusions. Some 
results regarding the spectral representation of nucleon 
three-point functions are collected in an appendix.

\section{Axial Charge Violation in the Field of an Instanton}
\label{sec_anom}

 We would like to start by showing explicitly how the axial
anomaly is realized in the field of an instanton. This 
discussion will be useful for the calculation of the OZI
violating part of the axial-vector correlation function 
and the axial charge of the nucleon. The flavor singlet 
axial-vector current in a gluon background is given by    
\be 
\label{axcur}
A_\mu(x) = {\rm Tr}\left[\gamma_5\gamma_\mu S(x,x) \right]
\ee
where $S(x,y)$ is the full quark propagator in the background 
field. The expression on the right hand side of equ.~(\ref{axcur})
is singular and needs to be defined more carefully. We will 
employ a gauge invariant point-splitting regularization
\be
\label{ax_reg}
 {\rm Tr}\left[\gf\gamma_\mu S(x,x)\right] 
 \equiv \lim_{\epsilon\to 0}
  {\rm Tr}\left[\gf\gamma_\mu 
    S(x+\epsilon,x-\epsilon)
    P\exp\left(-i\int_{x-\epsilon}^{x+\epsilon}A_\mu(x)dx\right)
    \right].
\ee
In the following we will consider an (anti) instanton in singular 
gauge. The gauge potential of an instanton of size $\rho$ and 
position $z=0$ is given by
\be 
 A_\mu^a =  \frac{2\rho^2}{x^2+\rho^2}
  \frac{x^\nu}{x^2}
  \,R^{ab}\bar\eta^b_{\mu\nu}.
\ee
Here, $\bar{\eta}^a_{\mu\nu}$ is the 't Hooft symbol and
$R^{ab}$ characterizes the color orientation of the instanton.
The fermion propagator in a general gauge potential can be 
written as 
\be
\label{prop_sum}
S(x,y)=\sum_{\lambda}\frac{\Psi_\lb(x)\Psi_\lb^{+}(y)}{\lb-m},
\ee
where $\Psi_\lb(x)$ is a normalized eigenvector of the Dirac 
operator with eigenvalue $\lambda$, $\Dslash \Psi_\lb(x)=\lb
\Psi_\lb(x)$. We will consider the limit of small quark masses. 
Expanding equ.~(\ref{prop_sum}) in powers of $m$ gives
\be
\label{prop_expan}
S_{\pm}(x,y)=-\frac{\Psi_0(x)\Psi_0^{+}(y)}{m} 
  +  \sum_{\lb\ne 0}\frac{\Psi_\lb(x)\Psi_\lb^{+}(y)}{\lb}
  + m\sum_{\lb\ne 0}\frac{\Psi_\lb(x)\Psi_\lb^{+}(y)}{\lb^2} 
  + O(m^2).
\ee
Here we have explicitly isolated the zero mode propagator. 
The zero mode $\Psi_0$ was found by 't Hooft and is given by
\be 
\Psi_0(x) = \frac{\rho}{\pi} \frac{1}{(x^2+\rho^2)^{3/2}}
  \frac{\gamma\cdot x}{\sqrt{x^2}} \gamma_\pm \phi.
\ee
Here, $\phi^{a\alpha}=\epsilon^{a\alpha}/\sqrt{2}$ is a constant 
spinor and $\gpm=(1\pm\gf)/2$ for an instanton/anti-instanton. 
The second term in equ.~(\ref{prop_expan}) is the non-zero mode 
part of the propagator in the limit $m\to 0$ \cite{Brown:1977eb} 
\be
\label{vec_prop}
 S_{\pm}^{NZ}(x,y) 
   \equiv  \sum_{\lb\ne 0}\frac{\Psi_\lb(x)\Psi_\lb^{+}(y)}{\lb}
   =  \vec{\Dslash}_x\Delta_{\pm}(x,y)\gpm + 
              \Delta_\pm(x,y)\ivec{\Dslash}_y \gmp
\ee
where $D^\mu=\partial^\mu -iA_\pm^\mu$ and $\Delta_{\pm}(x,y)$ is 
the propagator of a scalar field in the fundamental representation.
Equ.~(\ref{vec_prop}) can be verified by checking that $S_{\pm}^{NZ}$
satisfies the equation of motion and is orthogonal to the zero mode.
The scalar propagator can be found explicitly
\be
\label{scal_prop}
\Delta_{\pm}(x,y)  =  \Delta_0(x,y)
   \frac{1}{\sqrt{1+\frac{\rho^2}{x^2}}}
   \left(1+\frac{\rho^2\sigma_{\mp}\cdot x
    \sigma_{\pm}\cdot y} {x^2 y^2}\right)
   \frac{1}{\sqrt{1+\frac{\rho^2}{y^2}}} ,
\ee
where $\Delta_0=1/(4\pi^2\Delta^2)$ with $\Delta = x-y$
is the free scalar propagator. The explicit form of the 
non-zero mode propagator can be obtained by substituting 
equ.~(\ref{scal_prop}) into equ.~(\ref{vec_prop}). We find 
\bea\label{NZ_prop}
 S^{NZ}_{\pm}(x,y) &=&
  \frac{1}{\sqrt{1+\frac{\rho^2}{x^2}}}
  \frac{1}{\sqrt{1+\frac{\rho^2}{y^2}}} 
  \left\{ S_0(x-y)\left(1+\frac{\rho^2\sigma_\mp\cdot x\sigma_{\pm}\cdot y} 
      {x^2y^2}\right) \right. \nonumber \\
 & & \hspace{0.3cm}  \mbox{}
  -  \frac{\Delta_0(x-y)}{x^2y^2}
     \left(\frac{\rho^2}{\rho^2+x^2}\sigma_\mp\cdot x 
         \sigma_\pm\cdot \gamma
         \sigma_\mp\cdot \Delta \sigma_\pm\cdot y \gpm 
    \right. \nonumber \\
 & & \hspace{2.5cm}  \mbox{}\left.\left.
  + \frac{\rho^2}{\rho^2+y^2}\sigma_\mp\cdot x \sigma_\pm\cdot 
      \Delta \sigma_\mp\cdot \gamma \sigma_\pm\cdot y \gmp
 \right) \right\}
\eea
Here, $S_0=-\slashed{\Delta}/(2\pi^2\Delta^4)$ denotes the free 
quark propagator. As expected, the full non-zero mode
propagator reduces to the free propagator at short distance. The 
linear mass term in equ. (\ref{prop_expan}) can be written in 
terms of the non-zero mode propagator 
\be
\label{int_SS}
 \sum_{\lb\ne 0}\frac{\Psi_\lb(x)\Psi_{\lb}^+(y)}{\lb^2}=
  \int d^4z S^{NZ}_\pm(x,z) S^{NZ}_\pm(z,y) 
  =-\Delta_\pm(x,y)\gpm -\Delta^M_\pm(x,y) \gmp ,
\ee
where $\Delta_\pm(x,y)$ is the scalar propagator and $\Delta^M_\pm
(x,y)=\langle x|(D^2+\sigma\cdot G/2)^{-1}|y\rangle$ is the 
propagator of a scalar particle with a chromomagnetic moment.
We will not need the explicit form of $\Delta^M_\pm(x,y)$
in what follows. We are now in the position to compute the 
regularized axial current given in equ.~(\ref{ax_reg}). We 
observe that neither the free propagator nor the zero mode 
part will contribute. Expanding the non-zero mode propagator 
and the path ordered exponential in powers of $\epsilon$ we find
\be
\label{ax_cur_inst}
{\rm Tr}\left[\gf\gamma^\mu S(x,x)\right]
 =\pm\frac{2\rho^2x^\mu}{\pi^2(x^2+\rho^2)^3},
\ee
which shows that instantons act as sources and sinks for the 
flavor singlet axial current. We can now compare this result 
to the anomaly relation equ.~(\ref{anom}). The divergence of 
equ.~(\ref{ax_cur_inst}) is given by 
\be 
\partial^\mu A_\mu(x) = \pm
  \frac{2\rho^2}{\pi^2}\frac{4\rho^2-2x^2}{(x^2+\rho^2)^4}.
\ee
The topological charge density in the field of an 
(anti) instanton is 
\be 
\label{qtop_i}
\frac{g^2}{16\pi^2}G^a_{\mu\nu} \tilde{G}^a_{\mu\nu}
 = \pm\frac{12\rho^4}{\pi^2(x^2+\rho^2)^4}.
\ee
We observe that the divergence of the axial current given in 
equ.~(\ref{ax_cur_inst}) does not agree with the topological 
charge density. The reason is that in the field of an instanton 
the second term in the anomaly relation, which is proportional 
to $m\bar{\psi}\gamma_5 \psi$, receives a zero mode contribution 
and is enhanced by a factor $1/m$. In the field of an (anti) 
instanton we find
\be 
\label{mq5q}
2m\bar{\psi}i\gamma_5 \psi = \mp
 \frac{4\rho^2}{\pi^2(x^2+\rho^2)^3}.
\ee
Taking into account both equ.~(\ref{qtop_i}) and (\ref{mq5q})
we find that  the anomaly relation (\ref{anom}) is indeed 
satisfied. 

\section{OZI Violation in Axial-Vector Two-Point Functions}
\label{sec_pia}

\begin{figure}
\begin{center}
\includegraphics[width=15cm,angle=0,clip=true]{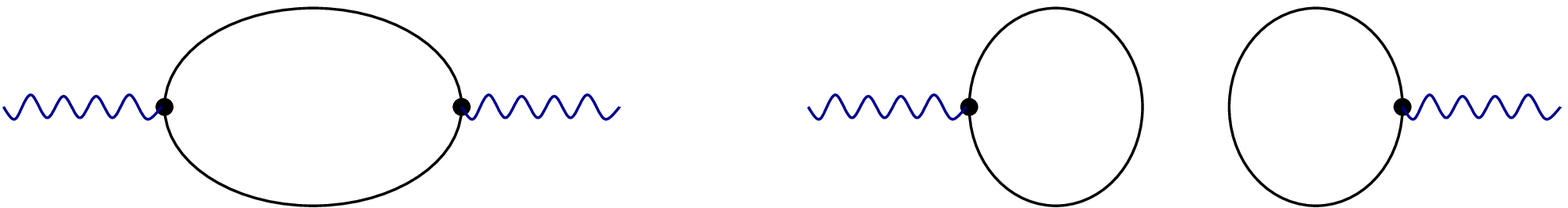}
\end{center}
\caption{\label{fig_vec_cor}
Quark line diagrams that contribute to the 
vector and axial-vector two-point function in the 
iso-vector and iso-singlet channel. The solid lines
denote quark propagators in a gluonic background field.
The two diagrams show the connected and disconnected 
contribution.}
\end{figure}

 In this section we wish to study OZI violation in the axial-vector 
channel due to instantons. We consider the correlation functions 
\be 
\Pi_{\mu\nu}^{j}(x,y) =\langle j_\mu(x)j_\nu(y)\rangle ,
\ee
where $j_\mu$ is one of the currents
\be 
\begin{array}{rcllrcll}
V_\mu^a &=& \bar{\psi}\gamma_\mu\tau^a\psi & (\rho), \hspace{1.5cm}&
V_\mu^0 &=& \bar{\psi}\gamma_\mu\psi & (\omega), \\
A_\mu^a &=& \bar{\psi}\gamma_\mu\gamma_5\tau^a\psi & (a_1), 
\hspace{1.5cm}&
A_\mu^0 &=& \bar{\psi}\gamma_\mu\gamma_5\psi & (f_1), 
\end{array}
\ee
where in the brackets we have indicated the mesons with the 
corresponding quantum numbers. We will work in the chiral limit 
$m_u=m_d\to 0$. The iso-vector correlation functions only 
receive contributions from connected diagrams. The iso-vector
vector ($\rho$) correlation function is 
\be
\label{triplet}
(\Pi_V^3)_{\mu\nu}(x,y) = 2(P^{con}_V)_{\mu\nu}(x,y)
 = -2 \langle {\rm Tr}\left[ 
  \gamma_\mu S(x,y)\gamma_\nu S(y,x)\right]\rangle .
\ee
The iso-singlet correlator receives additional, disconnected, 
contributions, see Fig.~\ref{fig_vec_cor}. The iso-singlet 
vector ($\omega$) correlator is given by
\be
\label{singlet}
(\Pi_V^0)_{\mu\nu}(x,y)  = 2(P^{con}_V)_{\mu\nu}(x,y) 
  + 4(P^{dis}_V)_{\mu\nu}(x,y)
\ee
with 
\be
(P^{dis}_V)_{\mu\nu}(x,y) = \langle 
  {\rm Tr}\left[\gamma_\mu S(x,x)\right] 
  {\rm Tr}\left[\gamma_\nu S(y,y)\right] \rangle .
\ee
The axial-vector correlation functions are defined analogously. 
At very short distance the correlation functions are dominated 
by the free quark contribution $\Pi_A^3=\Pi_A^0=\Pi_V^3=\Pi_V^0
\sim 1/x^6$. Perturbative corrections to the connected correlators
are $O(\alpha_s(x)/\pi)$, but perturbative corrections to the 
disconnected correlators are very small, $O((\alpha_s(x)/\pi)^2)$. 
In this section we will compute the instanton contribution 
to the correlation functions. At short distance, it is 
sufficient to consider a single instanton. For the connected 
correlation functions, this calculation was first performed
by Andrei and Gross \cite{Andrei:xg}, see also \cite{Nason:1993ak}.
Disconnected correlation function were first considered in 
\cite{Geshkenbein:vb} and a more recent study can be found
in \cite{Dorokhov:2003kf}.

 In order to make contact with our calculation of the vector 
and axial-vector three-point functions in the next section, we 
briefly review the calculation of Andrei and Gross, and then 
compute the disconnected contribution. Using the expansion in 
powers of the quark mass, equ.~(\ref{prop_expan}), we can write 
\be
\label{vec_dec}
 (P^{con}_V)_\pm^{\mu\nu}(x,y)
  = (P^{con}_V)^{\mu\nu}_{0} + A^{\mu\nu}_{\pm} + B^{\mu\nu}_\pm 
\ee
with 
\bea 
(P_V^{con})^{\mu\nu}_{0} &=&  
  -{\rm Tr}\left[\gamma^{\mu}S_0(x,y)\gamma^{\nu}S_0(y,x)\right],\\
\label{A}
A^{\mu\nu}_\pm  &=&
  -{\rm Tr}\left[\gamma^{\mu}S^{NZ}_\pm(x,y)
                 \gamma^{\nu}S^{NZ}_\pm(y,x)\right]
  - (P^{con}_V)_0^{\mu\nu}(x,y), \\
\label{B}
B^{\mu\nu}_\pm  &=&
  -2 {\rm Tr}\left[\gamma^\mu\Psi_{0\pm}(x)\Psi_{0\pm}^+(y)
                   \gamma^\nu \Delta_{\pm}(y,x)\gpm\right].
\eea
Using the explicit expression for the propagators given in 
the previous section we find
\bea
\label{A_2}
A^{\mu\nu}_{\pm}&=&
  \frac{\rho^2 h_x h_y}{2\pi^4\Delta^4}
  \left\{
    S^{\mu\alpha\nu\beta}
    \Big[\rho^2 h_x h_y(2\Delta_\alpha\Delta_\beta - 
      \Delta^2\delta_{\alpha\beta}) +
      h_y(y_\beta \Delta_\alpha + y_\alpha \Delta_\beta) 
  \right. \nonumber \\
 & & \hspace{2cm}\mbox{} \left. 
   - h_x (x_\beta 
      \Delta_\alpha + x_\alpha \Delta_\beta) \Big]
     \mp 2\epsilon^{\mu\nu\alpha\beta}(h_y\Delta_\alpha y_\beta 
  - h_x   \Delta_\beta x_\alpha)
  \right\}
\eea
and 
\be
\label{B_2}
B^{\mu\nu}_\pm = -\frac{\rho^2}{\pi^4\Delta^2}h_x^2 h_y^2
 \left\{
   (x\cdot y +\rho^2)\delta^{\mu\nu} - (y^\mu x^\nu - x^\mu y^\nu) 
   \mp \epsilon^{\mu\alpha\nu\beta}x_\alpha y_\beta
 \right\}
\ee
with $h_x=1/(x^2+\rho^2)$, $\Delta=x-y$, and $S^{\mu\alpha\nu\beta}
=g^{\mu\alpha}g^{\nu\beta} - g^{\mu\nu}g^{\alpha\beta} + g^{\mu\beta}
g^{\alpha\nu}$. Our result agrees with \cite{Andrei:xg} up to a color 
factor of $3/2$, first noticed in \cite{Dubovikov:bf}, a '-' sign in 
front of the epsilon terms, which cancels after adding instantons and 
anti-instantons, and a '-' sign in front of the 2nd term in $B^{\mu\nu}$. 
This sign is important in order to have a conserved current, but it
does not affect the trace $P^{\mu\mu}$. Summing up the contributions 
from instantons and anti-instantons we obtain
\bea
(P^{con}_V)^{\mu\nu} (x,y)&=&
 2\frac{12 S^{\mu\alpha\nu\beta} \Delta_\alpha 
    \Delta_\beta}{(2\pi^2)^2\Delta^8}
  + \frac{1}{2\pi^4}(h_x h_y)^2\rho^2 
    \left[
     -\frac{2\Sigma^2}{\Delta^4} \Delta^\mu\Delta^\nu
     +\frac{2\Sigma\cdot\Delta}{\Delta^4}  
      (\Sigma^\mu\Delta^\nu + \Delta^\mu 
       \Sigma^\nu  \right. \nonumber \\
 & & \hspace{2cm} \mbox{}\left.
   - \Sigma\cdot \Delta g^{\mu\nu})
   + \frac{2}{\Delta^2}  (\Delta^2g^{\mu\nu} - \Delta^\mu\Delta^\nu - 
   \Delta^\mu\Sigma^\nu + \Delta^\nu \Sigma^\mu)
   \right] .
\eea 
This result has to be averaged over the position of the instanton.
We find
\bea
 2a^{\mu\nu} &=& \sum_\pm \int d^4 z A^{\mu\nu}_\pm (x-z,y-z) 
 = -\frac{1}{\pi^2} \left[ 
    \frac{\partial^2}{\partial\Delta_\mu\partial\Delta_\nu} 
    G(\Delta^2,\rho) + 2G'(\Delta^2,\rho)g^{\mu\nu} \right], \\
 2b^{\mu\nu} &=& \sum_\pm \int d^4 z B^{\mu\nu}_\pm (x-z,y-z) 
 =\frac{1}{\pi^2}\left[ 
   \frac{\partial^2}{\partial\Delta^2} 
    G(\Delta^2,\rho) + 2G'(\Delta^2,\rho)\right]g^{\mu\nu},
\eea
with 
\be
 \label{G}
 G'(\Delta^2,\rho) = 
  \frac{\partial G(\Delta^2,\rho)}{\partial \Delta^2}
  = \frac{\rho^2}{\Delta^4}
    \left[-\frac{2\rho^2}{\Delta^2}\xi
        \log\frac{1-\xi}{1+\xi}-1 \right]
\ee
and $\xi^2= \Delta^2/(\Delta^2+4\rho^2)$. The final result 
for the single instanton contribution to the connected part 
of the vector current correlation function is
\be
\delta (P_V^{con})^{\mu\mu}= 
      (P_V^{con})^{\mu\mu} - 2 (P^{con}_V)^{\mu\mu}_0
  =\frac{24}{\pi^2}\frac{\rho^4}{\Delta^2} \;
  \frac{\partial}{\partial \Delta^2}
  \left(\frac{\xi}{\Delta^2}\log\frac{1+\xi}{1-\xi}
  \right)
  \equiv \frac{24}{\pi^2}\frac{\rho^4}{\Delta^2} F(\Delta,\xi),
\ee
where we defined 
\be
F(\Delta,\xi) = \frac{\partial}{\partial \Delta^2}
\left(\frac{\xi}{\Delta^2}\log\frac{1+\xi}{1-\xi}
\right) .
\ee
The computation of the connected part of the axial-vector 
correlator is very similar. Using equs.~(\ref{A},\ref{B})
we observe that the only difference is the sign in front 
of $B^{\mu\nu}$. We find
\be
\delta (P^{con}_A)^{\mu\mu}= 
  (P^{con}_A)^{\mu\mu} - 2 (P^{con}_A)^{\mu\mu}_{0} 
=-\frac{1}{\pi^2}\left[ 20\Delta^2G'' + 56 G' \right],
\ee
with $G'$ given in equ.~(\ref{G})

\begin{figure}
\begin{center}
\leavevmode
\includegraphics[width=16cm,angle=0,clip=true]{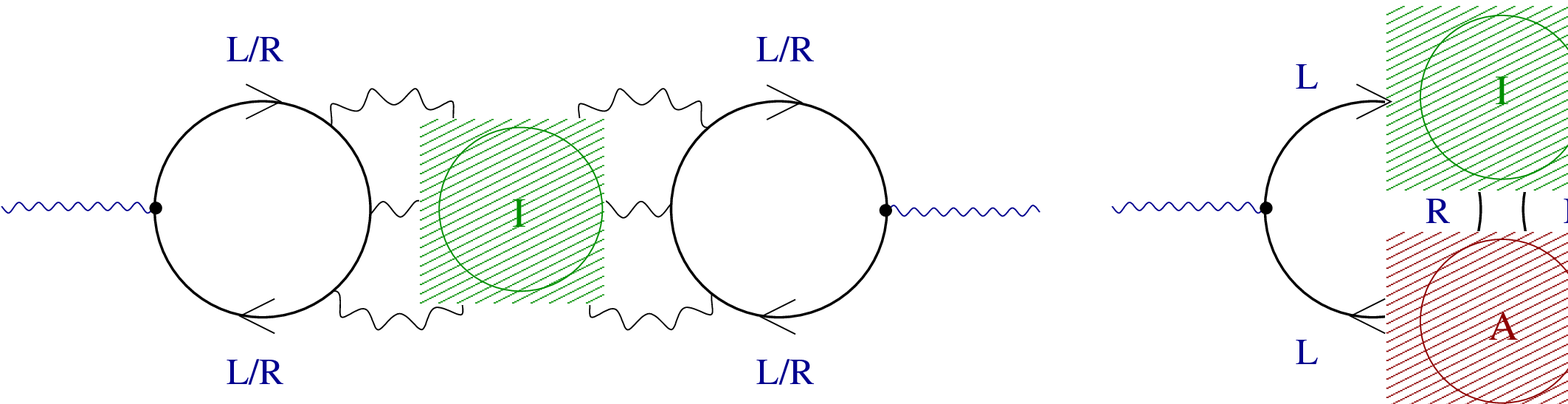}
\end{center}
\caption{\label{fig_f1_inst}
Instanton contributions to the disconnected axial-vector
correlation function. The left panel shows the single-instanton
(non-zero mode) contribution. The right panel shows the 
instanton-anti-instanton (fermion zero mode) contribution.}
\end{figure}

\begin{figure}
\begin{center}
\includegraphics[width=9cm,angle=0,clip=true]{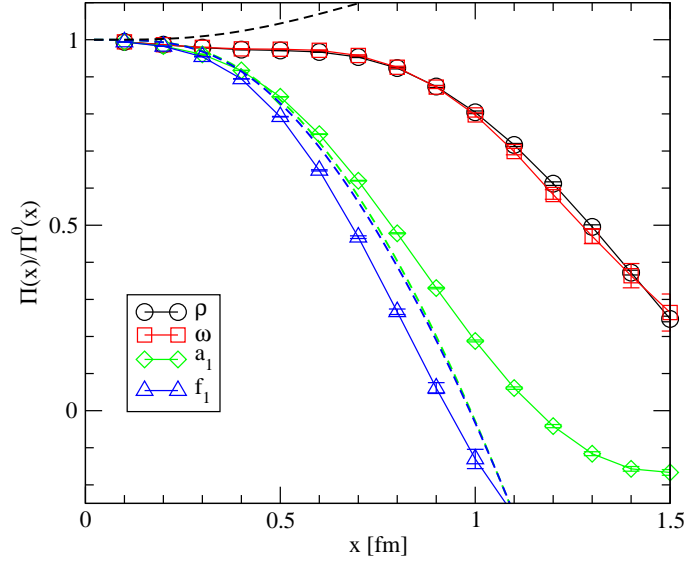}
\end{center}
\caption{\label{fig_vec_cor_num}
Correlation functions in the $\rho,\omega,a_1$ and
$f_1$ channel. All correlation functions are normalized 
to free field behavior. The data points show results from 
a numerical simulation of the random instanton liquid. The 
dashed lines show the single instanton approximation.}
\end{figure}

 We now come to the disconnected part, see Fig.~\ref{fig_f1_inst}. 
In the vector channel the single instanton contribution to 
the disconnected correlator vanishes \cite{Geshkenbein:vb}. 
In the axial-vector channel we can use the result for 
${\rm Tr}[\gf\gamma^\mu S(x,x)]$ derived in the previous 
section. The correlation function is
\be
(P^{dis}_A)^{\mu\nu}(x,y)=\frac{4\rho^4(x-z)^\mu(y-z)^\nu}
 {\pi^4((x-z)^2+\rho^2)^3((y-z)^2+\rho^2)^3} .
\ee
Summing over instantons and anti-instantons and integrating 
over the center of the instanton gives
\be
(P^{dis}_A)^{\mu\nu} = 2\frac{\rho^4}{2\pi^2}
 \frac{\partial^2}{\partial \Delta_\mu\partial \Delta_\nu}
   F(\Delta,\xi)
\ee
and 
\be
 (P^{dis}_A)^{\mu\mu} = \frac{4\rho^4}{\pi^2}
 \left\{2\frac{d}{d\Delta^2} + 
       \Delta^2\left(\frac{d}{d\Delta^2}\right)^2\right\}
        F(\Delta,\xi). 
\ee
We can now summarize the results in the vector singlet 
($\omega$) and triplet ($\rho$), as well as axial-vector
singlet ($f_1$) and triplet ($a_1$) channel. The result 
in the $\rho$ and $\omega$ channel is 
\be
(\Pi^3_V)^{\mu\mu} = (\Pi^0_V)^{\mu\mu}=
 -\frac{12}{\pi^4\Delta^6} +
  2\int d\rho n(\rho) \frac{24}{\pi^2}\frac{\rho^4}{\Delta^2}
  F(\Delta,\xi) .
\ee
In the $a_1,f_1$ channel we have
\bea
(\Pi^3_A)^{\mu\mu} &=& 
   -\frac{12}{\pi^4\Delta^6} +
   \int d\rho n(\rho) \left[ - \frac{2}{\pi^2}
     \left(20\Delta^2G'' + 56 G'\right)  \right], \\
(\Pi^0_A)^{\mu\mu} &=& 
 -\frac{12}{\pi^4\Delta^6} +
 \int d\rho n(\rho) 
   \left[ - \frac{2}{\pi^2}\left(20\Delta^2G'' + 56 G'\right)
  + \frac{16\rho^4}{\pi^2} \left( 2F'+\Delta^2F''\right)  \right].
\eea
In order to obtain a numerical estimate of the instanton 
contribution we use a very simple model for the instanton
size distribution, $n(\rho)=n_0\delta(\rho-\bar{\rho})$,
with $\bar{\rho}=0.3$ fm and $n_0=0.5\, {\rm fm}^{-4}$. The 
results are shown in Fig.~\ref{fig_vec_cor_num}. 

 We observe that the OZI rule violating difference 
between the singlet and triplet axial-vector correlation 
functions is very small and repulsive. We can also see
this by studying the short distance behavior of the 
correlation functions. The non-singlet correlators 
satisfy 
\bea
(\Pi_V^3)^{\mu\mu} &=& (\Pi_V^3)^{\mu\mu}_{0} 
  \left\{ 1 + \frac{\pi^2x^4}{3} 
  \int d\rho n(\rho) + \ldots \right\}, \\
(\Pi_A^3)^{\mu\mu} &=& (\Pi_A^3)^{\mu\mu}_{0} 
  \left\{ 1 - \pi^2x^4
  \int d\rho n(\rho) + \ldots \right\}.
\eea
As explained by Dubovikov and Smilga, this result 
can be understood in terms of the contribution of the 
dimension $d=4$ operators $\langle g^2G^2\rangle$ and 
$\langle m\bar{q}q\rangle$ in the operator product 
expansion (OPE). The OZI violating contribution 
\be
(\Pi^{OZI}_A)^{\mu\mu} =  
(\Pi^0_A)^{\mu\mu} - (\Pi^3_A)^{\mu\mu}
 =  -(\Pi^3_A)^{\mu\mu}_0
   \left(\frac{4\pi^2}{45}\frac{x^6}{\rho^2}\right)
  \int d\rho n(\rho)
\ee
is of $O(x^6)$ and not singular at short distance. Our 
results show that it remains small and repulsive even if 
$x>\rho$. We also note that the sign of the OZI-violating 
term at short distance is model independent. The quark 
propagator in euclidean space satisfies the Weingarten 
relation
\be
\label{wein}
S(x,y)^\dagger  = \gamma_5 S(y,x) \gamma_5 .
\ee
This relation implies that ${\rm Tr}[S(x,x)\gamma_\mu\gamma_5]$ 
is purely real. As a consequence we have
\be 
\label{ozi_ineq}
\lim_{x\to y}  \Big\{ {\rm Tr}[S(x,x)\gamma_\mu\gamma_5]
  {\rm Tr}[S(y,y)\gamma_\mu\gamma_5] \Big\} > 0.
\ee
Since the path integral measure in euclidean space is 
positive this inequality translates into an inequality
for the correlation functions. In our convention the 
trace of the free correlation function is negative, and
equ.~(\ref{ozi_ineq}) implies that the interaction is 
repulsive at short distance. The result is in agreement 
with the single instanton calculation. 

 We can also study higher order corrections to the single 
instanton result. The two-instanton (anti-instanton) 
contributions is of the same form as the one-instanton
result. An interesting contribution arises from 
instanton-anti-instanton pairs, see Fig.~\ref{fig_f1_inst}. 
This effect was studied in \cite{Schafer:1994nv}. It was 
shown that the instanton-anti-instanton contribution to 
the disconnected meson channels can be described in terms 
of an effective lagrangian 
\be 
{\cal L} = \frac{2G}{N_c^2}(\bar{\psi}\gamma_\mu\gamma_5\psi)^2
\ee
with 
\be 
G = \int d\rho_1d\rho_2 (2\pi\rho_1)^2(2\pi\rho_2)^2
 \frac{n(\rho_1,\rho_2)}{8T_{IA}^2}
\ee
where $n(\rho_1,\rho_2)$ is the tunneling rate for an 
instanton-anti-instanton pair and $T_{IA}$ is the 
matrix element of the Dirac between the two (approximate)
zero modes. We note that this interaction is also repulsive,
and that there is no contribution to the $\omega$ channel. 
  
\begin{figure}
\begin{center}
\vspace{1cm}
\includegraphics[width=9cm,angle=0,clip=true]{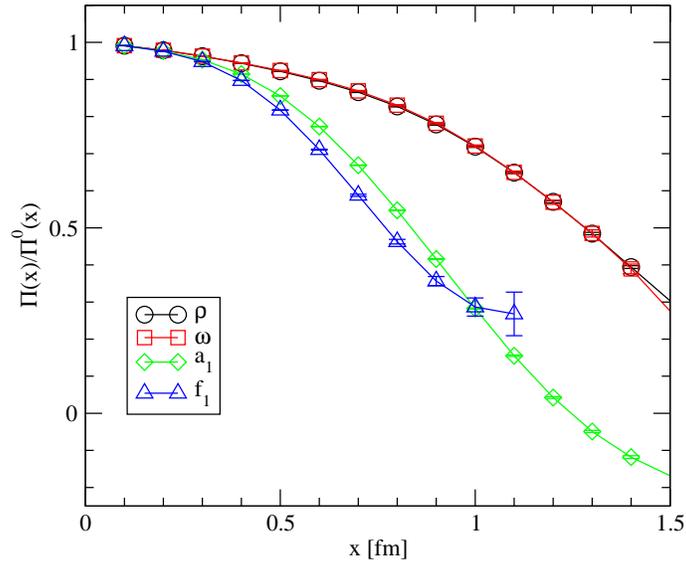}
\end{center}
\caption{\label{fig_vec_cor_unq}
Correlation functions in the $\rho,\omega,a_1$ and
$f_1$ channel. All correlation functions are normalized 
to free field behavior. The data points show results from 
unquenched simulations of the instanton liquid model.}
\end{figure}

 Numerical results for the vector meson correlation functions 
are shown in Figs.~\ref{fig_vec_cor_num} and \ref{fig_vec_cor_unq}.
The correlation functions are obtained from Monte Carlo 
simulations of the instanton liquid as described in 
\cite{Schafer:1995pz,Schafer:1995uz}. We observe that OZI 
violation in the vector channel is extremely small, both in 
quenched and unquenched simulations. The OZI violating 
contribution to the $f_1$ channel is repulsive. In quenched 
simulations this contribution becomes sizable at large distance. 
Most likely this is due to mixing with an $\eta'$ ghost pole. 
We observe that the effect disappears in unquenched simulations. 
The pion contribution to the $a_1$ correlator is of course
present in both quenched and unquenched simulations.

 Experimentally we know that the $\rho$ and $\omega$, as 
well as the $a_1$ and $f_1$ meson, are indeed almost 
degenerate. Both iso-singlet states are slightly heavier
than their iso-vector partners. To the best of our knowledge
there has been only one attempt to measure OZI violating 
correlation functions in the vector and axial-vector channel 
on the lattice, see \cite{Isgur:2000ts}. Isgur and Thacker
concluded that OZI violation in both channels was too 
small to be reliably measurable in their simulation.

\section{Axial Vector Coupling of a Quark}
\label{sec_gaq}

 In this section we wish to study the iso-vector and iso-singlet 
axial coupling of a single quark. Our purpose is twofold. One
reason is that the calculation of the axial-vector three-point 
function involving a single quark is much simpler than that of 
the nucleon, and that it is closely connected to the axial-vector 
two-point function studied in the previous section. The second, 
more important, reason is the success of the constituent quark 
model in describing many properties of the nucleon. It is clear 
that constituent quarks have an intrinsic structure, and that the 
axial decay constant of a constituent quark need not be close to 
one. Indeed, Weinberg argued that the axial coupling of a quark 
is $(g_{A}^3)_Q \simeq 0.8$ \cite{Weinberg:gf}. Using this value 
of $(g_{A}^3)_Q$ together with the naive $SU(6)$ wave function of 
the nucleon gives the nucleon axial coupling $g_A^3=0.8\cdot 5/3\simeq 
1.3$, which is a significant improvement over the naive quark model
result $5/3$. It is interesting to study whether, in a similar fashion, 
the suppression of the flavor singlet axial charge takes place on the 
level of a constituent quark. 

\begin{figure}
\begin{center}
\includegraphics[width=6.5cm,angle=0,clip=true]{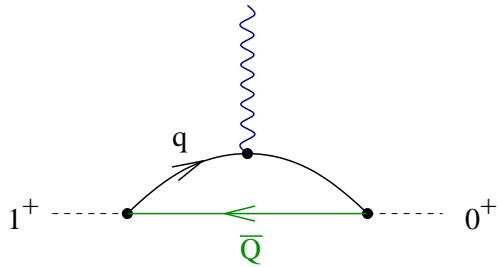}
\end{center}
\caption{\label{fig_hl}
Physical interpretation of the gauge invariant 
axial-vector three-point function of a quark in 
terms of a weak light-quark transition in a heavy-light
$\bar{Q}q$ meson.}
\end{figure}

 In order to address this question we study three-point functions
involving both singlet and triplet vector and axial-vector 
currents. The vector three-point function is
\be
\label{VQQ}
(\Pi^a_{VQQ})^{\alpha\beta}_\mu(x,z,y) = 
\langle q^\alpha(x)V^a_\mu(z)\bar{q}^\beta(y)\rangle .
\ee
The axial-vector function $(\Pi^a_{AQQ})$ is defined analogously.
We should note that equ.~(\ref{VQQ}) is not gauge invariant.
We can define a gauge invariant correlation function by 
including a gauge string. The gauge string can be interpreted
as the propagator of a heavy anti-quark, see Fig.~\ref{fig_hl}. 
This implies that the gauge invariant quark axial-vector 
three-point function is related to light quark weak transitions 
in heavy-light mesons. 
 
\begin{figure}
\begin{center}
\leavevmode
\includegraphics[width=12cm,angle=0,clip=true]{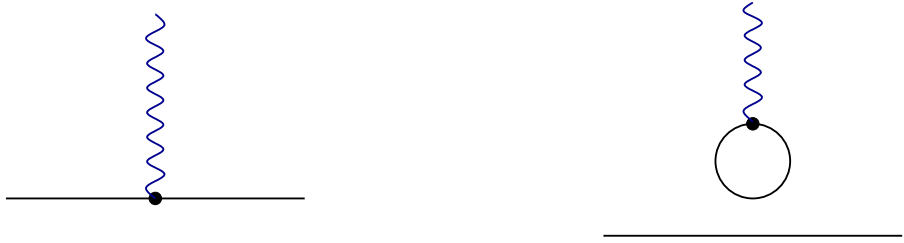}
\end{center}
\caption{\label{fig_VQQ}
Quark line diagrams that contribute to the 
vector and axial-vector three-point function of a
constituent quark. The solid lines denote quark 
propagators in a gluonic background field.
The two diagrams show the connected and disconnected 
contribution.}
\end{figure}

  The spectral representation of vector and axial-vector 
three-point functions is studied in some detail in the 
appendix. The main result is that in the limit that 
$y_4\gg z_4 \gg x_4$ the ratio 
\be
  \frac{{\rm Tr}[(\Pi^a_{AQQ})_3 \gamma_5\gamma_3]}
                {{\rm Tr}[(\Pi^a_{VQQ})_4 \gamma_4]}
 \to \frac{g_A}{g_V}
\ee
tends to the ratio of axial-vector and vector coupling
constants, $g_A/g_V$. We therefore define the following 
Dirac traces
\bea
\label{correl_def}
(\Pi^{a}_{VQQ})^{\mu\nu}(x,z,y)
  &=& {\rm Tr}[(\Pi^{a}_{VQQ})^{\mu} \gamma^\nu], \\
(\Pi^{a}_{AQQ})^{\mu\nu}(x,z,y)
  &=& {\rm Tr}[(\Pi^{a}_{AQQ})^{\mu} \gf \gamma^\nu] .
\label{correl_ax_def} 
\eea
As in the case of the two-point function the iso-triplet 
correlator only receives quark-line connected contributions,
whereas the iso-singlet correlation function has a disconnected 
contribution, see Fig.~\ref{fig_VQQ}. We find
\be 
\label{Pi_tripl}
(\Pi^3_{VQQ})^{\mu\nu} (x,z,y) =  (P^{con}_{VQQ})^{\mu\nu}(x,z,y) 
 = \langle {\rm Tr} [S(x,z)\gamma^\mu  S(z,y)\gamma^\nu] \rangle 
\ee
and 
\be
\label{Pi_singl}
(\Pi^0_{VQQ})^{\mu\nu}(x,z,y) = (P^{con}_{VQQ})^{\mu\nu}(x,z,y)
   -2(P^{dis}_{VQQ})^{\mu\nu}(x,z,y)
\ee
with
\be
(P^{dis}_{VQQ})^{\mu\nu}(x,z,y) = 
 \langle {\rm Tr} [S(x,y)\gamma^\mu]{\rm Tr}[S(z,z)\gamma^\nu]
 \rangle ,
\ee
as well as the analogous result for the axial-vector correlator.

 In the following we compute the single instanton contribution
to these correlation functions. We begin with the connected 
part. We again write the propagator in the field of the 
instanton as $S_{ZM}+S_{NZ}+S_m$ where $S_{ZM}$ is the 
zero-mode term, $S_{NZ}$ is the non-zero mode term, and
$S_m$ is the mass correction. In the three-point correlation
function we get contribution of the type $S_{NZ}S_{NZ}$, 
$S_{ZM}S_{m}$ and $S_{m}S_{ZM}$
\bea
\label{pi_3pt_z/nz}
 (P^{con}_{A/VQQ})^{\mu\nu} &=& 
  P_{NZNZ}^{\mu\nu} + c_{A/V}
   \left(P^{\mu\nu}_{ZMm} + P^{\mu\nu}_{mZM}
   \right)\nonumber\\
  &=& {\rm Tr} [S^{NZ}(x,z)\gamma^{\mu}S^{NZ}(z,y)\gamma^{\nu}]
\nonumber\\ 
& & +
c_{A/V}{\rm Tr} [-\Psi_0(x)\Psi_0^{+}(z)\gamma^{\mu}
(-\Delta_{\pm}(z,y)\gamma_{\pm}\gamma^{\nu}]\nonumber\\
& & +
c_{A/V }{\rm Tr} [(-\Delta_{\pm}(x,z)\gamma_{\pm})\gamma^{\mu}
  (-\Psi_0(z)\Psi_0^{+}(y))\gamma^{\nu}] ,
\eea
where the only difference between the vector and axial-vector 
case is the sign of 'ZMm' and 'mZM' terms. We have  $c_{A/V}
=\pm 1$ for vector (axial vector) current insertions. The 
detailed evaluation of the traces is quite tedious and we 
relegate the results to the appendix \ref{sec_app5}.

\begin{figure}
\begin{center}
\includegraphics[width=12cm,angle=0,clip=true]{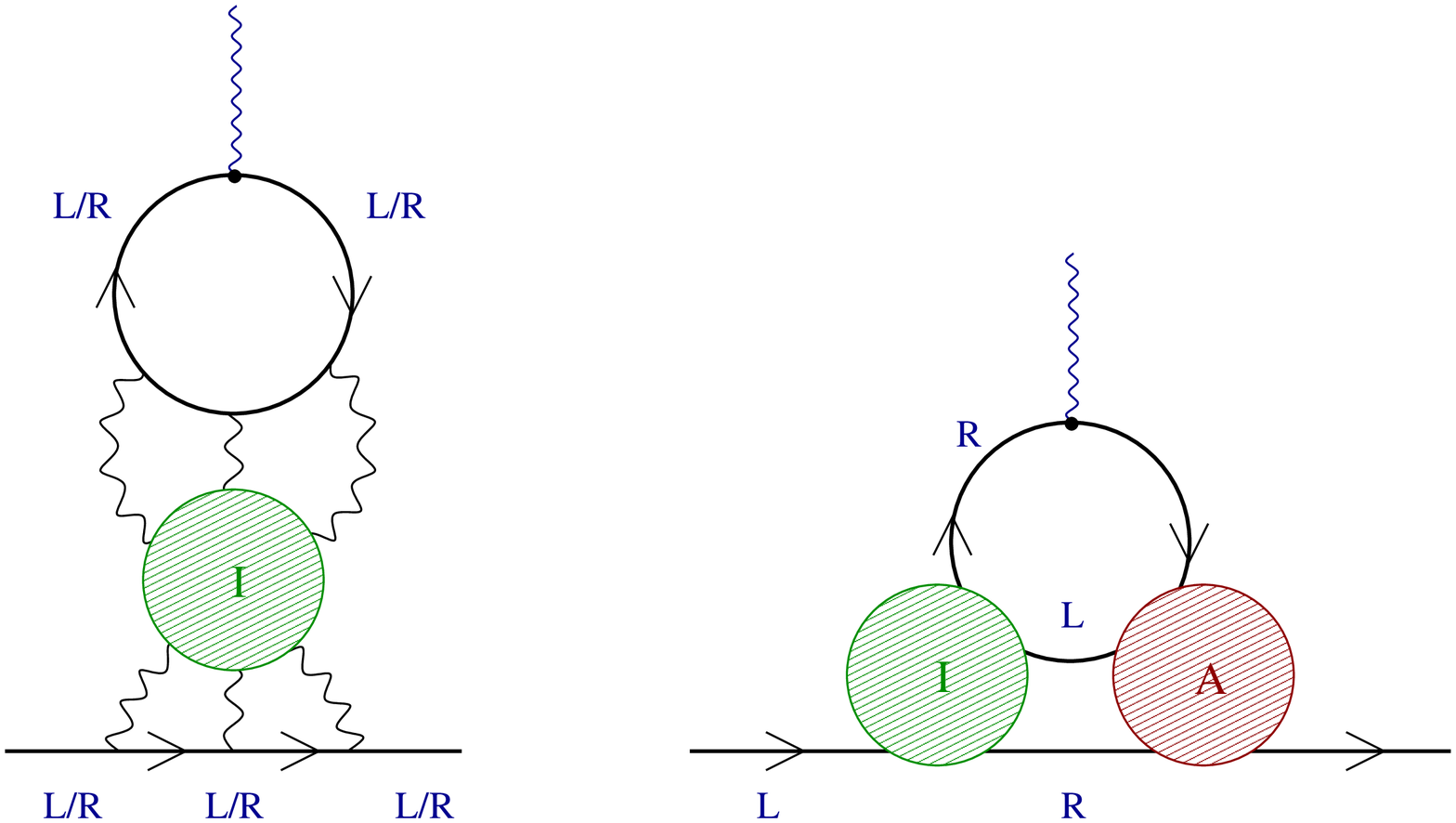}
\end{center}
\caption{\label{fig_gaq_inst}
Instanton contributions to the disconnected axial-vector
three-point correlation function of a quark. The left panel 
shows the single-instanton (non-zero mode) contribution. The 
right panel shows the instanton-anti-instanton (fermion zero 
mode) contribution.}
\end{figure}

 Our main goal is the calculation of the disconnected 
correlation function, which is related to OZI violation. 
In the single instanton approximation only the axial-vector
correlator receives a non-zero disconnected contribution
\be
(P^{dis}_{AQQ})^{\mu\nu}={\rm Tr}[S(x,y)\gf\gamma^\nu] 
  {\rm Tr}[S(z,z)\gf \gamma^\mu], 
\ee
see Fig.~\ref{fig_gaq_inst}. We observe that the second trace 
is the axial-vector current in the field of an instanton, see 
equ.~(\ref{ax_cur_inst}). As for the first trace, it is easy 
to see that neither the zero mode part of the propagator nor 
the part of $S_{NZ}$ proportional to the free propagator can
contribute. A straight-forward computation gives
\be
{\rm Tr} [S(x,y)\gf\gamma^\nu]=\mp\frac{\rho^2}{\pi^2(x-y)^2x^2y^2}
 \frac{x^\alpha y^\beta (x-y)^\sigma}
 {\sqrt{(1+\frac{\rho^2}{x^2})(1+\frac{\rho^2}{y^2})}}
 \left(\frac{S^{\alpha\nu\sigma\beta}}{\rho^2+x^2} -
       \frac{S^{\alpha\sigma\nu\beta}}{\rho^2+y^2} \right).
\ee
Combined with equ.~(\ref{ax_cur_inst}) we obtain
\be
(P^{dis}_{AQQ})^{\mu\nu}= - 
  \frac{2\rho^4 z^\mu x^\alpha y^\beta (x-y)^\sigma}
  {\pi^4(x-y)^2x^2y^2(z^2 + \rho^2)^3}
   \frac{1}{\sqrt{(1+\frac{\rho^2}{x^2})(1+\frac{\rho^2}{y^2})}}
  \left(\frac{S^{\alpha\nu\sigma\beta}}{\rho^2+x^2} -
        \frac{S^{\alpha\sigma\nu\beta}}{\rho^2+y^2}\right),
\ee
which has to be multiplied by a factor 2 in order to take 
into account both instantons and anti-instantons. 

\begin{figure}
\begin{center}
\leavevmode
\includegraphics[width=9cm,angle=0,clip=true]{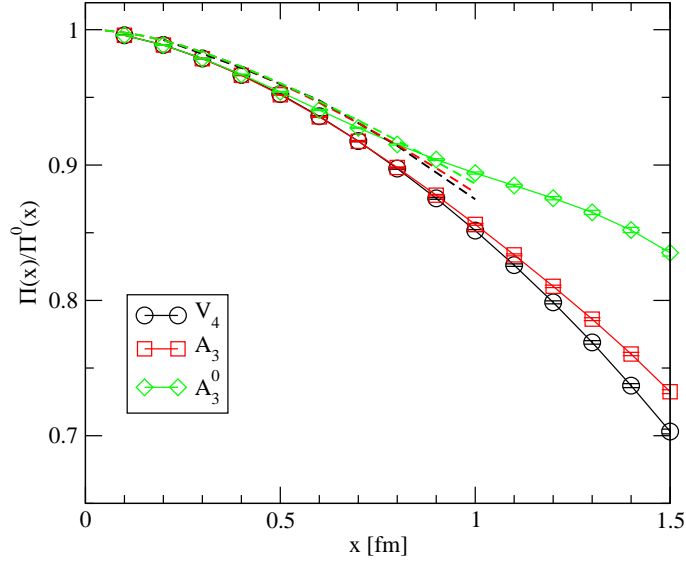}
\end{center}
\caption{\label{fig_qu_va}
Axial and vector three-point functions of a quark as a 
function of the separation between the two quark sources.
The correlation functions are normalized to free field 
behavior. The data points show results from numerical 
simulations of the instanton liquid and the dashed lines
show the single instanton approximation.}
\end{figure}

  Results for the vector and axial-vector three-point functions 
$(\Pi_{VQQ}^3)^{44}(\tau,\tau/2,0)$ and $(\Pi_{AQQ}^{0,3})^{33}
(\tau,\tau/2,0)$ are shown in Fig.~\ref{fig_qu_va}. We observe 
that the vector and axial-vector correlation functions are very 
close to one another. We also note that the disconnected
contribution adds to the connected part of the axial-vector 
three-point function. This can be understood from the short 
distance behavior of the correlation function. The disconnected 
part of the gauge invariant three-point function satisfies
\be 
\lim_{y,z\to x} \Big\{ (\Pi_{AQQ}^{dis})^{33}(x,z,y) \Big\}
 = \lim_{x\to z} \Big\{ {\rm Tr}[S(x,x)\gamma_3\gamma_5]
  {\rm Tr}[S(z,z)\gamma_3\gamma_5] \Big\} >0.
\ee
This expression is exactly equal to the short distance term in the 
disconnected $f_1$ meson correlation function. The short distance 
behavior of the connected three-point function, on the other hand, 
is opposite in sign to the two-point function. This is related to the 
fact that the two-point function involves one propagator in the forward 
direction and one in the backward direction, whereas the three-point 
function involves two forward propagating quarks. A similar connection 
between the interaction in the $f_1$ meson channel and the flavor 
singlet coupling of a constituent quark was found in a Nambu-Jona-Lasinio 
model \cite{Vogl:1991qt,Steininger:ed}. It was observed, in particular, 
that an attractive coupling in the $f_1$ channel is needed in 
order to suppress the flavor singlet $(g_A^0)_Q$.   

\begin{figure}
\begin{center}
\leavevmode
\includegraphics[width=9cm,angle=0,clip=true]{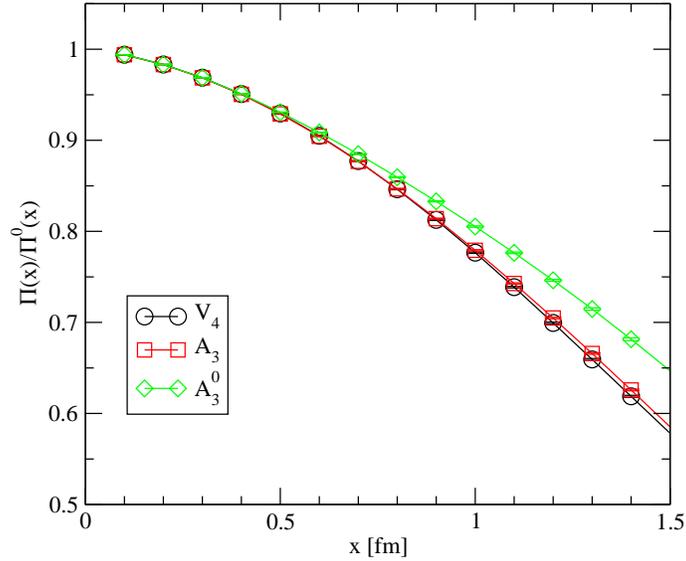}
\end{center}
\caption{\label{fig_qu_va_unq}
Axial and vector three-point functions of a quark 
as a function of the separation between the two quark
sources. The data points show results from an unquenched
instanton simulation.}
\end{figure}

\begin{figure}
\begin{center}
\leavevmode
\vspace{1cm}
\includegraphics[width=10cm,angle=0,clip=true]{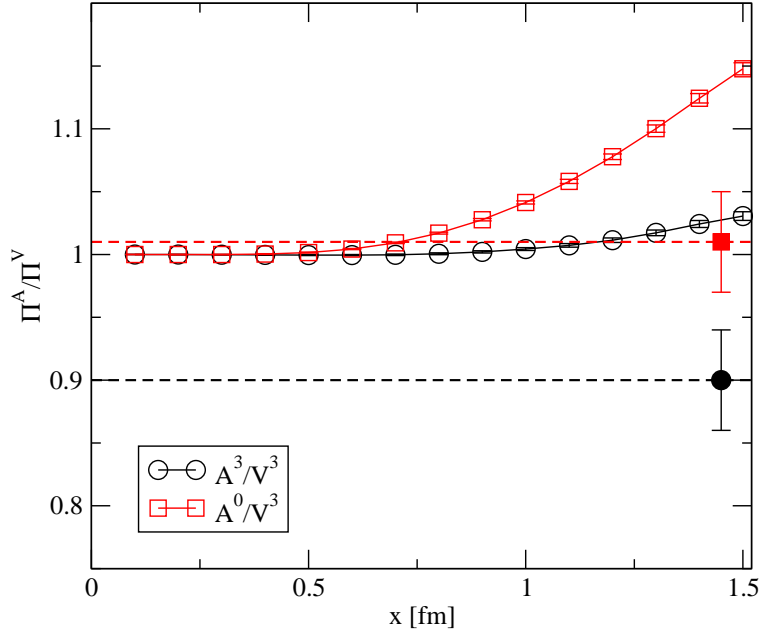}
\end{center}
\caption{\label{fig_qu_ga}
Ratio of axial-vector to vector correlation functions 
of a constituent quark calculated in the instanton 
liquid model. The open points show point-to-point
correlation functions while the solid point is the 
zero momentum (point-to-plane) limit. The figure 
shows the iso-vector and iso-singlet correlation functions.}
\end{figure}

 The same general arguments apply to the short distance contribution 
from instanton-anti-instanton pairs. At long distance, on the other 
hand, we expect that IA pairs reduce the flavor singlet axial current 
correlation function. The idea can be understood from Fig.~\ref{fig_gaq_inst}, 
see \cite{Dorokhov:ym,Dorokhov:2001pz,Kochelev:1997ux}. In an IA 
transition a left-handed valence up quark emits a right handed 
down quark which acts to shield its axial charge. We have studied 
this problem numerically, see Figs.~\ref{fig_qu_va}-\ref{fig_qu_ga}.
We find that in quenched simulations the flavor singlet axial
three-point function is significantly enhanced. This effect is 
analogous to what we observed in the $f_1$ channel and disappears
in unquenched simulations. We have also studied the axial three-point 
function at zero three-momentum $\vec{q}=0$. This correlation function 
is directly related to the coupling constant, see App.~\ref{sec_app}. 
We find that the iso-vector coupling is smaller than one, $(g_A^3)_Q
\simeq 0.9$, in agreement with Weinberg's idea. The flavor singlet 
coupling, on the other hand, is close to one. We observe no suppression of 
the singlet charge of a constituent quark. We have also checked 
that this result remains unchanged in unquenched simulations. 

\section{Axial Structure of the Nucleon}
\label{sec_gan}
 
  In this section we shall study the axial charge of the 
nucleon in the instanton model. We consider the same
correlation functions as in the previous section, but 
with the quark field replaced by a nucleon current. The
vector three-point function is given by
\be
(\Pi^a_{VNN})^{\alpha\beta}_\mu(x,y) = 
\langle \eta^\alpha(0)V^a_\mu(y)\bar{\eta}^\beta(x)\rangle .
\ee
Here, $\eta^\alpha$ is a current with the quantum numbers
of the nucleon. Three-quark currents with the nucleon
quantum numbers were introduced by Ioffe \cite{Ioffe:kw}.
He showed that there are two independent currents with 
no derivatives and the minimum number of quark fields
that have positive parity and spin $1/2$. In the case 
of the proton, the two currents are
\bea
\label{ioffe}
\eta_1 = \epsilon_{abc} (u^a C\gamma_\mu u^b) 
            \gamma_5 \gamma_\mu d^c, \hspace{1cm}
\eta_2 = \epsilon_{abc} (u^a C\sigma_{\mu\nu} u^b) 
            \gamma_5 \sigma_{\mu\nu} d^c .
\eea
It is sometimes useful to rewrite these currents in terms of scalar
and pseudo-scalar diquark currents. We find
\bea
\label{ioffe_ps}
\eta_{1} &=& 2 \left\{ \epsilon_{abc} (u^a C d^b)\gamma_5 u^c 
  - \epsilon_{abc} (u^a C\gamma_5 d^b) u^c \right\}, \\
\eta_{2} &=& 4 \left\{ \epsilon_{abc} (u^a C d^b)\gamma_5 u^c 
  + \epsilon_{abc} (u^a C\gamma_5 d^b) u^c \right\}.
\eea
Instantons induce a strongly attractive interaction in the scalar 
diquark channel $\epsilon^{abc}(u^bC\gamma_5d^c)$ 
\cite{Schafer:1993ra,Rapp:1997zu}. As a consequence, the nucleon 
mainly couples to the scalar diquark component of the Ioffe currents 
$\eta_{1,2}$. This phenomenon was also observed on the lattice 
\cite{Leinweber:1994nm}. This result is suggestive of a model 
of the spin structure that is quite different from the naive 
quark model. In this picture the nucleon consists of a tightly 
bound scalar-isoscalar diquark, loosely coupled to the third 
quark \cite{Anselmino:1992vg}. The quark-diquark model suggests 
that the spin and isospin of the nucleon are mostly carried by a 
single constituent quark, and that $g_A^N\simeq g_A^Q$.

\begin{figure}
\begin{center}
\includegraphics[width=15cm,angle=0,clip=true]{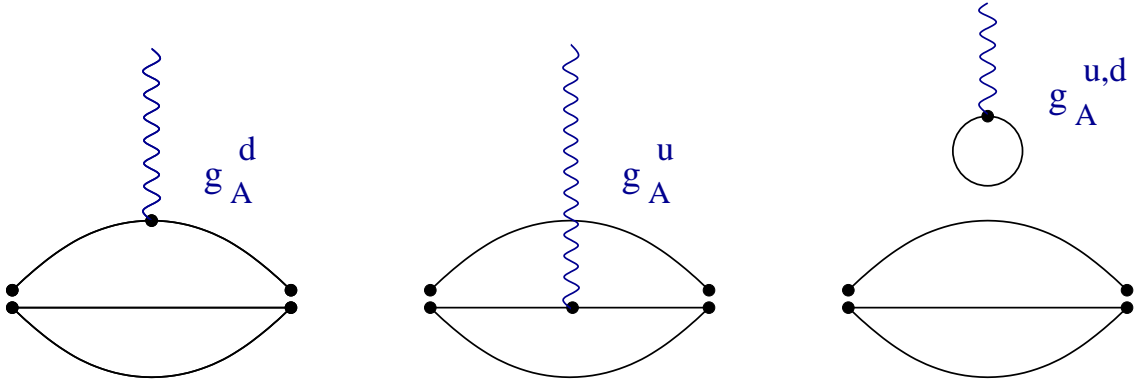}
\end{center}
\caption{\label{fig_VNN}
Quark line diagrams that contribute to the 
axial-vector three-point function of the proton. 
The solid lines denote quark propagators in a gluonic 
background field. The lines are connected in the same
way that the Dirac indices of the propagators are 
contracted. The iso-vector and iso-singlet correlation functions 
correspond to $g_A^u=-g_A^d=1$ and $g_A^u=g_A^d=1$, 
respectively. The disconnected diagram only contributes
to the iso-scalar three-point function.}
\vspace*{3cm}
\end{figure}

 Nucleon correlation functions are defined by $\Pi^N_{\alpha\beta}
(x) = \langle \eta_\alpha(0)\bar\eta_\beta(x) \rangle$, where 
$\alpha,\beta$ are Dirac indices. The correlation function of the 
first Ioffe current is
\be
\label{nn_cor}
 \Pi_{\alpha\beta}(x)=2\epsilon_{abc}\epsilon_{a'b'c'}
    \,\langle \left( \gamma_\mu\gamma_5S_d^{cc'}(0,x)
       \gamma_\nu\gamma_5 \right)_{\alpha\beta}\,
 {\rm Tr}\left[
     \gamma_\mu S_u^{aa'}(0,x)\gamma_\nu C(S_u^{bb'}(0,x))^TC
        \right]\rangle .
\ee
The vector and axial-vector three-point functions can be 
constructed in terms of vector and axial-vector insertions
into the quark propagator, 
\bea
\label{SV}
(\Gamma^V_\mu)_f^{aa'}(x,y) &=& 
          S_f^{ab}(0,y)\gamma_\mu S_f^{ba'}(y,x), \\
\label{SA}
(\Gamma^A_\mu)_f^{aa'}(x,y) &=& 
         S_f^{ab}(0,y)\gamma_\mu\gamma_5 S_f^{ba'}(y,x). 
\eea
The three-point function is given by all possible 
substitutions of equ.~(\ref{SV}) and (\ref{SA}) into
the two-point function. We have 
\be
\begin{array}{rclc}
(\Pi^a_{VNN})^{\alpha\beta}_\mu(x,y) =
 2\epsilon_{abc}\epsilon_{a'b'c'}\hspace{-3.2cm}  & & & \\
  \langle \Bigg\{ & 
    g_V^d \left( \gamma_\rho\gamma_5 (\Gamma^V_\mu)_d^{cc'}(x,y)
       \gamma_\sigma\gamma_5 \right)_{\alpha\beta} & 
 {\rm Tr}\left[
     \gamma_\rho S_u^{aa'}(0,x)\gamma_\sigma C(S_u^{bb'}(0,x))^TC
        \right]  &  \\
  + & 2g_V^u  \left( \gamma_\rho\gamma_5 S_d^{cc'}(0,x)
       \gamma_\sigma\gamma_5 \right)_{\alpha\beta} & 
 {\rm Tr}\left[
     \gamma_\rho (\Gamma^V_\mu)_u^{aa'}(x,y)\gamma_\sigma C(S_u^{bb'}(0,x))^TC
        \right] &  \\
 - &  \left( \gamma_\rho\gamma_5S_d^{cc'}(0,x)
       \gamma_\sigma\gamma_5 \right)_{\alpha\beta} & 
 {\rm Tr}\left[
     \gamma_\rho S_u^{aa'}(0,x)\gamma_\sigma C(S_u^{bb'}(0,x))^TC
        \right]   &  \\
 & & \label{VNN_con} \cdot\;
 {\rm Tr}\left[g_V^u\gamma_\mu S_u^{dd}(y,y)
              +g_V^d\gamma_\mu S_d^{dd}(y,y)\right]
  &  \Bigg\}\rangle ,
\end{array}
\ee
where the first term is the vector insertion 
into the $d$ quark propagator in the proton, the second
term is the insertion into the $uu$ diquark, and the third 
term is the disconnected contribution, see Fig.~\ref{fig_VNN}. 
The vector charges of the quarks are denoted by $g_V^f$. In 
the case of the iso-vector three-point function we have $g_V^u=1$,
$g_V^d=-1$ and in the iso-scalar case $g_V^u=g_V^d=1$.

\begin{figure}
\begin{center}
\leavevmode
\vspace{1cm}
\includegraphics[width=10cm,angle=0,clip=true]{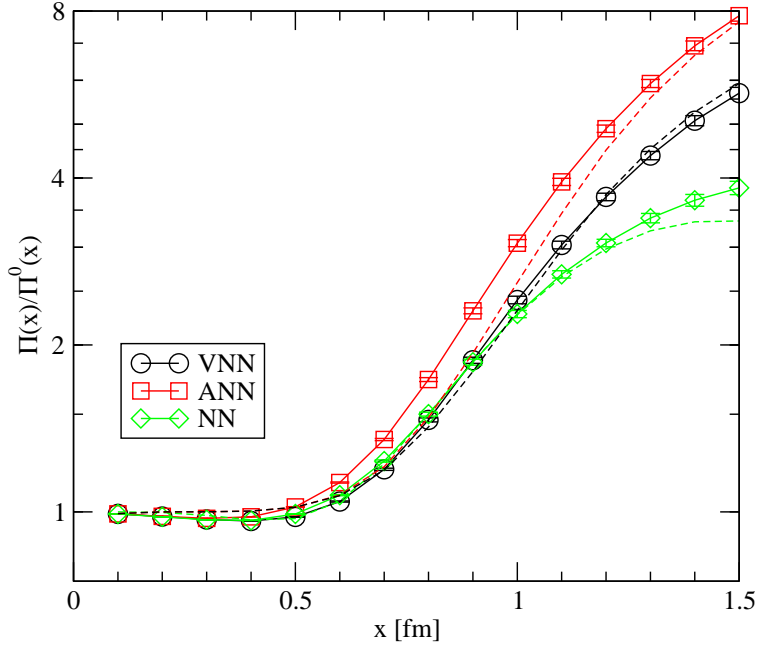}
\end{center}
\caption{\label{fig_nn_3pt}
Vector, axial-vector three-point functions 
of the nucleon and nucleon two-point function calculated
in the instanton liquid model. All correlation functions
are normalized to free field behavior. The results are 
compared to a simple pole fit of the type discussed 
in the appendix.}
\end{figure}

\begin{figure}
\begin{center}
\leavevmode
\includegraphics[width=9cm,angle=0,clip=true]{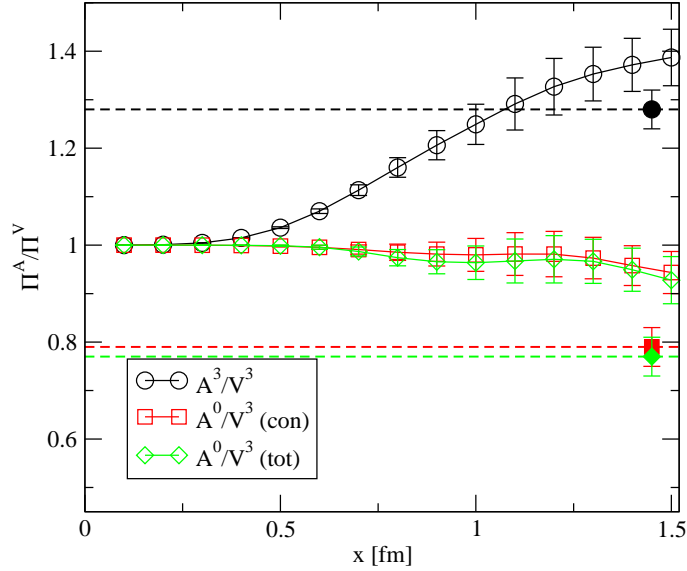}
\end{center}
\caption{\label{fig_gan}
Ratio of axial-vector to vector correlation
functions of the nucleon calculated in the instanton 
liquid model. The open points show point-to-point
correlation functions while the solid point is the 
zero momentum (point-to-plane) limit. The figure 
shows the iso-vector, connected iso-singlet, and 
full iso-singlet axial-vector correlation functions.}
\end{figure}

 Vector and axial-vector three-point functions of the nucleon
are shown in Figs.~\ref{fig_nn_3pt} and \ref{fig_gan}. In order to 
verify that the correlation functions are dominated by the nucleon 
pole contribution we have compared our results to the spectral 
representation discussed in the appendix, see Fig.~\ref{fig_nn_3pt}.
The nucleon coupling constant was determined from the nucleon
two-point function. The figure shows that we can describe 
the three-point functions using the phenomenological values
of the vector and axial-vector coupling constants. We have 
also checked that the ratio of axial-vector and vector 
current three-point functions is independent of the nucleon
interpolating field for $x>1$ fm. The only exception is 
a pure pseudo-scalar diquark current, which has essentially 
no overlap with the nucleon wave function. 

 The main result is that the iso-vector axial-vector correlation
function is larger than the vector correlator. The corresponding 
ratio is shown in Fig.~\ref{fig_gan}, together with the ratio 
of $\vec{q}=0$ correlation functions. We find that the iso-vector 
axial coupling constant is $g_A^3=1.28$, in good agreement with 
the experimental value. We also observe that the ratio of point-to-point
correlation functions is larger than this value. As explained in 
the appendix, this shows that the axial radius of the nucleon is 
smaller than the vector radius. Taking into account only the connected 
part of the correlation function we find a singlet coupling $g_A^0=
0.79$. The disconnected part is very small, $g_A^0(dis)=-(0.02
\pm 0.02)$. Assuming that $\Delta s\simeq \Delta u(dis)=
\Delta d (dis)$ this implies that the OZI violating difference 
$g_A^8-g_A^0$ is small. This result does not change in going
from the quenched approximation to full QCD.

\begin{figure}
\begin{center}
\leavevmode
\includegraphics[width=9cm,angle=0,clip=true]{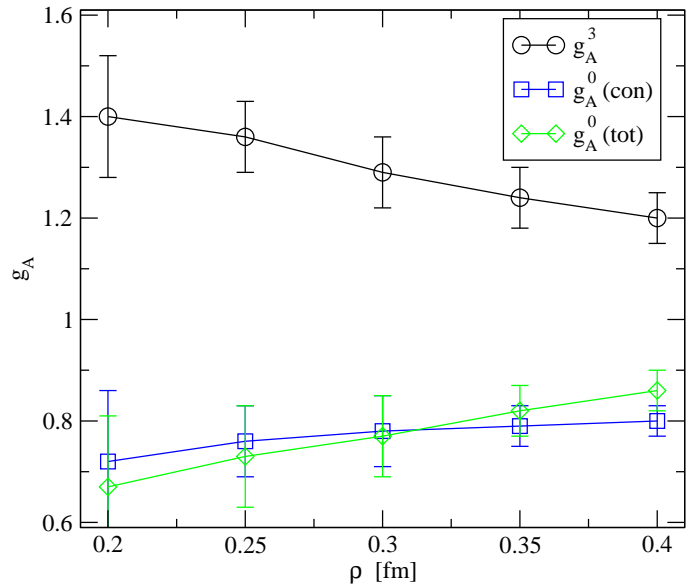}
\end{center}
\caption{\label{fig_gan_rho}
Axial coupling constants of the nucleon as a function of
the instanton size $\rho$ with the instanton density 
fixed at $(N/V)=1\,{\rm fm}^{-4}$. We show the iso-vector, 
connected iso-singlet, and full iso-singlet axial coupling
constant.}
\end{figure}

 We have also studied the dependence of the results on the 
average instanton size, see Fig.~\ref{fig_gan_rho}. We observe
that there is a slight decrease in the iso-singlet coupling
and a small increase in the iso-vector coupling as the instanton
size is decreased. What is surprising is that the disconnected
term changes sign between $\rho= 0.3$ fm and $\rho=0.35$ fm.
The small value $g_A^0(dis)=-(0.02\pm 0.02)$ obtained above 
is related to the fact that the phenomenological value of 
the instanton size is close to the value where $g_A^0(dis)$ 
changes sign. However, even for $\rho$ as small as 0.2 fm
the disconnected contribution to the axial coupling $g_A^0(dis)
=-(0.05\pm 0.02)$ is smaller in magnitude than phenomenology 
requires.

\section{Conclusions}
\label{sec_sum}

 The main issue raised by the EMC measurement of the
flavor singlet axial coupling is not so much why $g_A^0$ 
is much smaller than one -  except for the naive quark model
there is no particular reason to expect $g_A^0$ to be 
close to one - but why the OZI violating observable $g_A^0
-g_A^8$ is large. Motivated by this question we have studied 
the contribution of instantons to OZI violation in the axial-vector 
channel. We considered the $f_1-a_1$ meson splitting, the flavor 
singlet and triplet axial coupling of a constituent quark, and the 
axial coupling constant of the nucleon. We found that instantons 
provide a short distance contribution which is repulsive in the 
$f_1$ meson channel and adds to the gauge invariant flavor singlet 
three-point function of a constituent quark. We showed that 
the sign of this term is fixed by positivity arguments. 

 We computed the axial coupling constants of the constituent 
quark and the nucleon using numerical simulations of the 
instanton liquid. We find that the iso-vector axial coupling 
constant of a constituent quark is $(g_A^3)_Q=0.9$ and that of 
a nucleon is $g_A^3=1.28$, in good agreement with experiment. 
The result is also in qualitative agreement with the 
constituent quark model relation $g_A^3=5/3\cdot (g_A^3)_Q$.
The flavor singlet coupling of quark is close to one, while 
that of a nucleon is suppressed, $g_A^0=0.77$. However, this
value is still significantly larger than the experimental 
value $g_A^0=(0.28-0.41)$. In addition to that, we find very
little OZI violation, $\Delta s\simeq \Delta u(dis) \simeq
-0.01$. We observed, however, that larger values 
of the disconnected contribution can be obtained if the 
average instanton size is smaller than the phenomenological
value of $\rho \simeq 1/3$ fm.

 There are many questions that remain to be addressed. In
order to understand what is missing in our calculation it
would clearly be useful to perform a systematic study of 
OZI violation in the axial-vector channel on the lattice. 
The main question is whether the small value of $g_A^0$ 
is a property of the nucleon, or whether large OZI violation 
is also seen in other channels. A study of the connected
contributions to the axial coupling constant in cooled 
as well as quenched quantum QCD configurations was performed 
in \cite{Dolgov:1998js}. These authors find $g_A^0(con)=\Delta 
u(con)+\Delta d (con) \simeq 0.6$ in both cooled and full 
configurations. The disconnected term was computed by 
Dong et al.~\cite{Dong:1995rx}. They find $\Delta u(dis)
+\Delta d(dis)\simeq -0.24$. 

 In the context of the instanton model it is important to 
study whether the results for $g_A^0$ obtained from the 
axial-vector current three-point function are consistent
with calculations of $g_A^0$ based on the matrix element 
of the topological charge density $G\tilde{G}$
\cite{Forte:1990xb,Hutter:1995cs,Kacir:1996qn,Diakonov:1995qy}. 
It would also be useful to further clarify the connection of 
the instanton liquid model to soliton models of the nucleon 
\cite{Diakonov:1987ty}. In soliton models the spin of the 
nucleon is mainly due to the collective rotation of the 
pion cloud, and a small value for $g_A^0$ is natural 
\cite{Brodsky:1988ip,Blotz:1993am}. The natural parameter
that can be used in order to study whether this picture 
is applicable is the number of colors, $N_c$. Unfortunately, 
a direct calculations of nucleon properties for $N_c>3$ 
would be quite involved. Finally it would be useful to study 
axial form factors of the nucleon. It would be interesting to 
see whether there is a significant difference between the 
iso-vector and iso-singlet axial radius of the nucleon. A 
similar study of the vector form factors was recently presented 
in \cite{Faccioli:2003yy}.

Acknowledgments: We would like to thank E.~Shuryak for 
useful discussions. This work was supported in part by 
US DOE grant DE-FG-88ER40388.

\appendix
\section{Spectral Representation}
\label{sec_app}
\subsection{Nucleon Two-Point Function}
\label{sec_app1}

 Consider the euclidean correlation function
\be
\Pi^{\alpha\beta}_N(x)= \langle \eta^\alpha(0)\bar{\eta}^\beta(x)\rangle,
\ee
where $\eta^\alpha(x)$ is a nucleon current and $\alpha$ is a 
Dirac index. We can write 
\be 
\Pi^{\alpha\beta}_N(x)=  \Pi_1(x) (\hat{x}\cdot\gamma)^{\alpha\beta}
 +\Pi_2(x)\delta^{\alpha\beta}.
\ee
The functions $\Pi_{1,2}(x)$ have spectral representations
\bea
\Pi_1(x) &=& \int_0^\infty ds\,\rho_1(s) D'(\sqrt{s},x),\\
\Pi_2(x) &=& \int_0^\infty ds\,\rho_2(s) D(\sqrt{s},x),
\eea
where $\rho_{1,2}(s)$ are spectral functions and
\bea
D(m,x) &=& \;\frac{m}{4\pi^2x}K_1(mx) ,\\
D'(m,x)&=& - \frac{m^2}{4\pi^2x} K_2(mx) ,
\eea
are the euclidean coordinate space propagator of a scalar 
particle with mass $m$ and its derivative with respect to
$x$. The contribution to the spectral function arising from 
a nucleon pole is 
\be
\rho_1(s) = |\lambda^2_N| \delta(s-m_N^2), \hspace{1cm}
\rho_2(s) = |\lambda^2_N| m_N\delta(s-m_N^2),
\ee
where $\lambda_N$ is the coupling of the nucleon to the current,
$\langle 0 |\eta|N(p)\rangle = \lambda_N u(p)$, and $m_N$ is the 
mass of the nucleon. It is often useful to consider the point-to-plane 
correlation function
\be
K^{\alpha\beta}_N(\tau) = \int d^3x\, \Pi^{\alpha\beta}_N(\tau,
 \vec{x}).
\ee
The integral over the transverse plane insures that all intermediate 
states have zero three-momentum. The nucleon pole contribution 
to the point-to-plane correlation function is 
\be
K^{\alpha\beta}_N(\tau) = \frac{1}{2}(1+\gamma_4)^{\alpha\beta}
 |\lambda_N|^2 \exp(-m_N\tau).
\ee

\subsection{Scalar Three-Point Functions}
\label{sec_app2}

 Next we consider three-point functions. Before we get to 
three-point functions of spinor and vector currents we consider 
a simpler case in which the spin structure is absent. We study 
the three-point function of two scalar fields $\phi$ and a scalar 
current $j$. We define
\be 
\label{3pt_scal}
\Pi(x,y) = \langle \phi(0) j(y)\phi(x) \rangle .
\ee
The spectral representation of the three-point function is 
complicated and in the following we will concentrate on the 
contribution from the lowest pole in the two-point function
of the field $\phi$. We define the coupling of this state
to the field $\phi$ and the current $j$ as
\bea
\langle 0|\phi(0)|\Phi(p)\rangle &=& \lambda, \\
\langle \Phi(p')|j(0)|\Phi(p)\rangle &=& F(q^2),
\eea
where $F(q^2)$ with $q=p-p'$ is the scalar form factor. The 
pole contribution to the three-point function is
\be
\label{3pts_spec}
\Pi(x,y) = \lambda^2 \int d^4z\, D(m,y+z)F(z)D(m,x-y-z),
\ee
where $D(m,x)$ is the scalar propagator and $F(z)$ is 
the Fourier transform of the form factor. In order to 
study the momentum space form factor directly it is 
convenient to integrate over the location of the endpoint 
in the transverse plane and Fourier transform with 
respect to the midpoint 
\be 
\int d^3x\,\int d^3y\, e^{iqy}
 \langle \phi(0) j(\tau/2,\vec{y})\phi(\tau,\vec{x}) \rangle 
 =\frac{\lambda^2}{(2m)^2}\exp(-m\tau)F(q^2).
\ee
The correlation function directly provides the form factor
for spacelike momenta. Maiani and Testa showed that there 
is no simple procedure to obtain the time-like form factor
from euclidean correlation functions \cite{Maiani:ca}.
 
 Form factors are often parametrized in terms of monopole, dipole, 
or monopole-dipole functions
\bea 
F_M(q^2)    &=& F_M(0)\frac{m_V^2}{Q^2+m_V^2}, \\
F_D(q^2)    &=& F_D(0)\left(\frac{m_V^2}{Q^2+m_V^2}\right)^2, \\
F_{MD}(q^2) &=& F_{MD}(0)\frac{m_1^2}{Q^2+m_1^2}
 \left(\frac{m_2^2}{Q^2+m_2^2}\right)^2 ,
\eea
with $Q^2=-q^2$. For these parametrization the Fourier transform 
to euclidean coordinate space can be performed analytically. We find
\bea
 F_M(x) &=& m_V^2 D(x,m_V) \\
 F_D(x) &=& m_V^2\left(
    -\frac{x}{2} D'(x,m_V)-D(x,m_V)\right)\\
 F_{MD}(x) &=& \frac{m_1^2m_2^4}{m_2^2-m_1^2}
 \left\{ \frac{1}{m_2^2-m_1^2} 
      \left( D(x,m_1)-D(x,m_2) \right)\right. \nonumber \\
 & & \left.\hspace{2cm}\mbox{}+ \frac{1}{m_2^2} 
  \left( \frac{x}{2} D'(x,m_2) + D(x,m_2) \right) \right\}.
\eea
We also consider three-point functions involving a vector 
current $j_\mu$. The matrix element is 
\be
\langle \Phi(p')|j_\mu(0)|\Phi(p)\rangle = q_\mu F(q^2).
\ee
The pole contribution to the vector current three-point 
function is 
\be
\label{3pts_v_spec}
\Pi_\mu(x,y) = \lambda^2 \int d^4z\, D(m,y+z)\hat{z}_\mu 
     F'(z)D(m,x-y-z),
\ee
with $F'(z)=dF(z)/dz$ and $\hat{z}_\mu=z_\mu/|z|$. For the 
parameterizations given above the derivative of the coordinate 
space form factor can be computed analytically. We get 
\bea
 F_M'(x) &=& m_V^2 D'(x,m_V) \\
 F_D'(x) &=& -\frac{m_V^4}{2} D(x,m_V) \\
 F_{MD}'(x) &=& \frac{m_1^2m_2^4}{m_2^2-m_1^2}
 \left\{ \frac{1}{m_2^2-m_1^2} 
      \left( D'(x,m_1)-D'(x,m_2) \right)\right. \nonumber \\
 & & \left.\hspace{2cm}\mbox{}
    + \frac{x}{2} D(x,m_2)  \right\}.
\eea

\subsection{Three-Point Functions involving nucleons and 
vector or axial-vector currents }
\label{sec_app3}

 Next we consider three-point functions of the nucleon involving vector 
and axial-vector currents. The vector three-point function is 
\be
(\Pi^a_{VNN})^{\alpha\beta}_\mu(x,y) = 
\langle \eta^\alpha(0)V^a_\mu(y)\bar{\eta}^\beta(x)\rangle .
\ee
The axial-vector three-point function is defined analogously. 
The nucleon pole contribution involves the nucleon coupling 
to the current $\eta$ and the nucleon matrix element of the 
vector and axial vector currents. The vector current matrix 
element is 
\be
\langle N(p')|V_\mu^a|N(p)\rangle = 
 \bar{u}(p')\left[ F_1(q^2)\gamma_\mu
 +\frac{i}{2M}F_2(q^2) \sigma_{\mu\nu}q^\nu \right]
  \frac{\tau^a}{2}u(p),
\ee
where the form factors $F_{1,2}$ are related to the 
electric and magnetic form factors via
\bea
G_E(q^2) &=& F_1(q^2) +\frac{q^2}{4M^2}F_2(q^2), \\
G_M(q^2) &=& F_1(q^2) + F_2(q^2).
\eea
The axial-vector current matrix element is 
\be
\langle N(p')|A_\mu^a|N(p)\rangle = 
 \bar{u}(p')\left[ G_A(q^2)\gamma_\mu
 +\frac{1}{2M}G_P(q^2)(p'-p)_\mu \right]\gamma_5
  \frac{\tau^a}{2}u(p)
\ee
where $G_{A,P}$ are the axial and induced pseudo-scalar
form factors. 

 We are interested in extracting the vector and 
axial-vector coupling constants $g_V=F_1(0)$ and
$g_A=G_A(0)$. In order to determine the vector coupling
$g_V$ we study the three-point function involving 
the four-component of the vector current in the 
euclidean time direction. For simplicity we take
$y=x/2$. We find that 
\be
(\Pi_{VNN})^{\alpha\beta}_4(x,x/2) = 
   \Pi_{VNN}^{1}(\tau) \delta^{\alpha\beta}
 + \Pi_{VNN}^{2}(\tau) (\gamma_4)^{\alpha\beta},
\ee
where $x_\mu=(\vec{0},\tau)$. The two independent 
structures $\Pi_{VNN}^{1,2}$ are given by 
\bea 
\Pi_{VNN}^{1}(\tau) &=& |\lambda_N|^2\int d^4y\,
 \left\{ \left[ 
     \frac{\tau+2y_4}{2x_1}m D'(x_1)D(x_2)
    +\frac{\tau-2y_4}{2x_2}m D(x_1)D'(x_2) \right]F_1(y) 
      \right. \nonumber \\
 & & \hspace{1.5cm}\mbox{} +\left.
     \frac{\tau|\vec{y}|}{x_1x_2}D'(x_1)D'(x_2)
     \frac{F'_2(y)}{2m} \right\}, \\
\Pi_{VNN}^{2}(\tau) &=&  |\lambda_N|^2\int d^4y\,
 \left\{  \left[
   \frac{\tau^2-4y_4^2+4\vec{y}^2}{4x_1x_2} D'(x_1)D'(x_2) 
     +m^2D(x_1)D(x_2)  \right]F_1(y) 
      \right. \nonumber \\
 & & \hspace{1.5cm}\mbox{} +\left.
     |\vec{y}| \left[ 
     \frac{m}{x_1}D'(x_1)D(x_2) 
    +\frac{m}{x_2}D'(x_2)D(x_1) 
       \right] \frac{F'_2(y)}{2m} \right\},
\eea
where $F_{1,2}(y)$ are the Fourier transforms of the 
the Dirac form factors $F_{1,2}(q^2)$, and we have 
defined $x_{1}=(\vec{y},\tau/2+y_4)$ and $x_2=
(-\vec{y},\tau/2-y_4)$. 

 In order to extract the axial-vector coupling we study 
three-point functions involving spatial components of the 
axial-vector current. We choose the three-component of the 
current and again take $y=x/2$ with $x_\mu=(\vec{0},\tau)$. 
We find
\be
(\Pi_{ANN})^{\alpha\beta}_3(x,x/2) = 
   \Pi_{ANN}^{1}(\tau) (\gamma_5)^{\alpha\beta}
 + \Pi_{ANN}^{2}(\tau) (\gamma_3\gamma_5)^{\alpha\beta}
 + \Pi_{ANN}^{3}(\tau) (\gamma_3\gamma_4\gamma_5)^{\alpha\beta},
\ee
with 
\bea
\Pi_{ANN}^{1}(\tau) &=& |\lambda_N|^2\int d^4y \,
    my_3\left[ 
     \frac{\tau+2y_4}{2x_1} D'(x_1)D(x_2)
    -\frac{\tau-2y_4}{2x_2} D(x_1)D'(x_2) \right]G_A(y),\;\;\; \\
\Pi_{ANN}^{2}(\tau) &=&  |\lambda_N|^2\int d^4y\,
 \left\{  \left[
   \frac{\tau^2+8y_3^2-4y^2}{4x_1x_2} D'(x_1)D'(x_2) 
     +m^2D(x_1)D(x_2)  \right]G_A(y) 
      \right. \nonumber \\
 & & \hspace{1.5cm}\mbox{} +\left.
     \frac{y_3^2}{|\vec{y}|} \left[ 
     \frac{m}{x_1}D'(x_1)D(x_2) 
    +\frac{m}{x_2}D'(x_2)D(x_1) 
       \right] \frac{G'_P(y)}{2m} \right\},
\eea
where $G_{A,P}(y)$ are the Fourier transforms of the 
nucleon axial and induced pseudo-scalar form factors.

 These results are quite complicated. The situation 
simplifies if we consider three-point functions 
in which we integrate all points over their location 
in the transverse plane. The vector three-point function
is 
\be
\int d^3x\int d^3y\, (\Pi_{VNN})^{\alpha\beta}_4
 (\tau,\vec{x};\tau/2,\vec{y}) = 
    \frac{g_V}{2}(1+\gamma_4)^{\alpha\beta}
    |\lambda_N|^2 \exp(-m_N\tau), 
\ee
where $g_V=F_1(0)$ is the vector coupling. Note that 
the three-point function of the spatial components 
of the current vanishes when integrated over the 
transverse plane. The axial-vector three-point function
is 
\be
\int d^3x\int d^3y\, (\Pi_{ANN})^{\alpha\beta}_3
 (\tau,\vec{x};\tau/2,\vec{y}) = \frac{g_A}{2}
    \left((1+\gamma_4)\gamma_3\gamma_5\right)^{\alpha\beta}
    |\lambda_N|^2 \exp(-m_N\tau), 
\ee
where $g_A=G_A(0)$ is the axial-vector coupling. In
the case of the axial-vector current the three-point 
function of the time component of the current vanishes
when integrated over the transverse plane. This is why 
we consider three-point function involving the spatial 
components of the axial current. 

\subsection{Phenomenology}
\label{sec_app4}

\begin{figure}
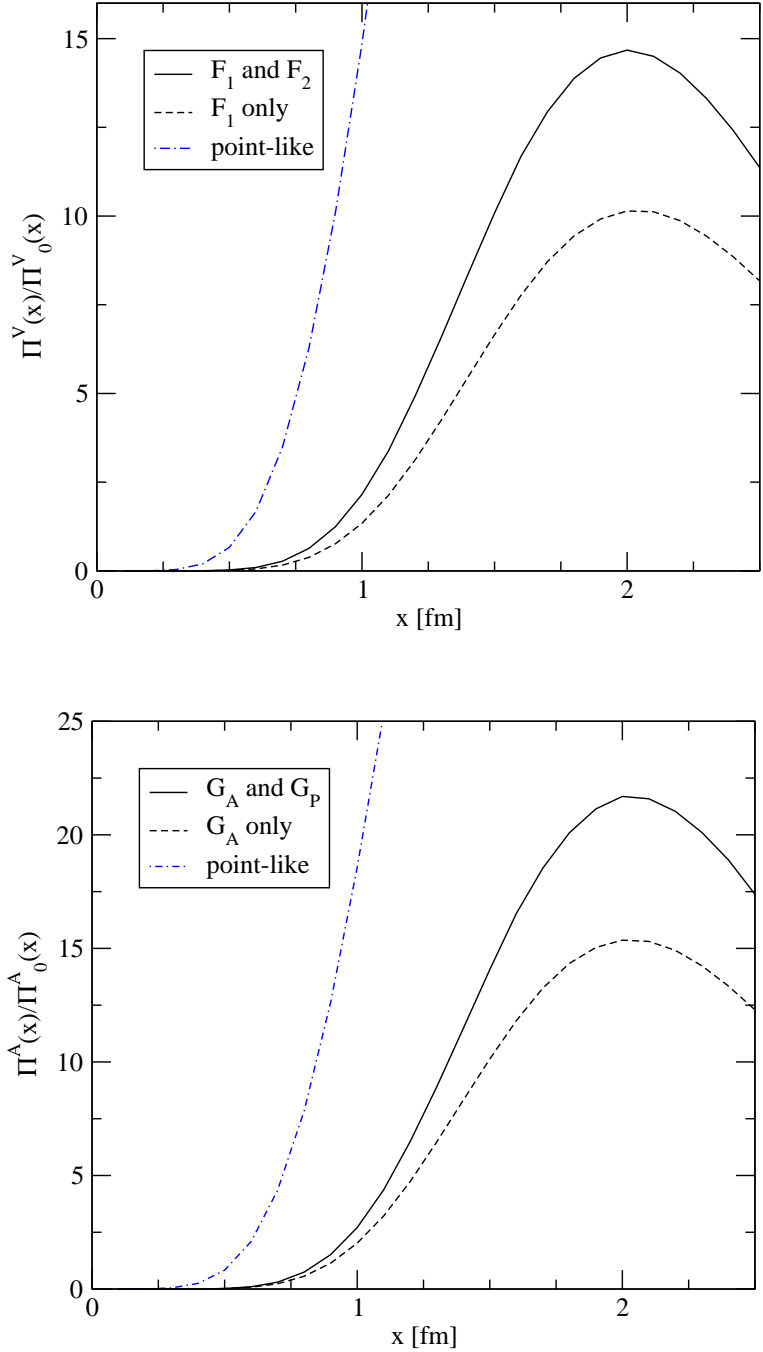

\begin{center}
\leavevmode
\vspace{1cm}
\includegraphics[width=10cm,angle=0,clip=true]{pheno_gv.eps}
\includegraphics[width=10cm,angle=0,clip=true]{pheno_ga.eps}
\end{center}
\caption{\label{fig_pheno}
Nucleon pole contribution to the vector (upper panel) and 
axial-vector (lower panel) nucleon three-point function.
The solid line shows the complete results, the dashed line
is the contribution from the $F_1$ and $G_A$ form factors
only, and the dash-dotted line corresponds to a point-like 
nucleon. We have used a nucleon coupling constant $\lambda=2.2\,
{\rm fm}^{-3}$ as well as phenomenological values for the 
form factors and coupling constants.}
\end{figure}

 In Fig.~\ref{fig_pheno} we show the nucleon pole contribution
to the vector and axial-vector three-point functions. We have
used the phenomenological values of the isovector coupling 
constants
\be
\begin{array}{rclcrcl}
 G_E(0) &=& 1,&\hspace{1cm} & G_M(0) &=& 4.7, \\
 G_A(0) &=& 1.25, & &         G_P(0) &=& \frac{4M^2}{m_\pi^2}g_A.
\end{array}
\ee 
We have parametrized $G_{E,V}$ and $G_A$ by dipole
functions with $m_V=0.88$ GeV  and $m_A=1.1$ GeV. The induced 
pseudoscalar form factor is parametrized as a pion propagator 
multiplied by a dipole form factor with dipole mass $m_A$. 

\begin{figure}
\begin{center}
\leavevmode
\vspace{1cm}
\includegraphics[width=10cm,angle=0,clip=true]{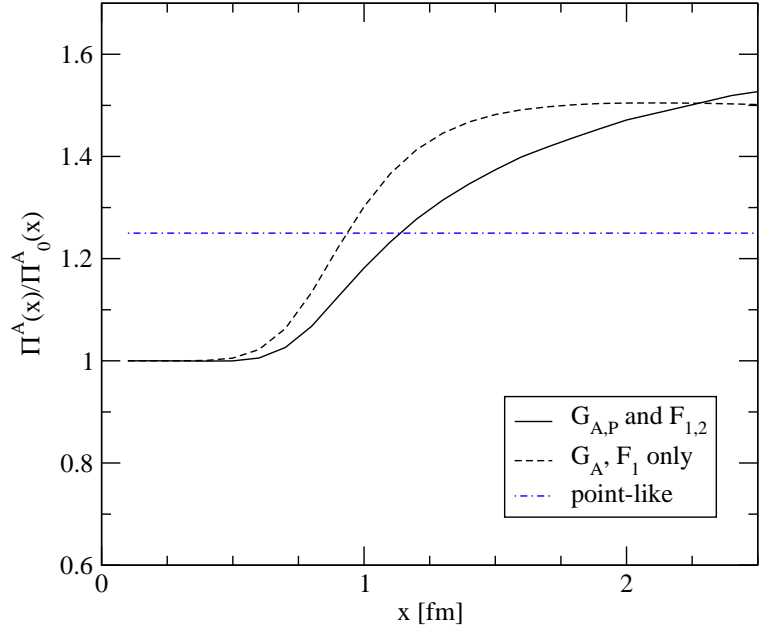}
\end{center}
\caption{\label{fig_pheno_rat}
Ratio of the phenomenological parameterizations of the 
axial-vector and vector three-point functions. We have 
added a short distance continuum contribution to the 
nucleon pole terms. The curves are labeled as in the 
previous figure. Note that both the solid and the dashed
line will approach $g_A=1.25$ as $x\to\infty$. Also 
note that the solid line is in very good agreement with 
the instanton calculation shown in Fig.~\ref{fig_gan}.}
\end{figure}

 We observe that at distances that are accessible in lattice 
or instanton simulations, $x\sim (1-2)$ fm, the typical momentum 
transfer is not small and the correlation function is substantially 
reduced as compared to the result for a point-like nucleon. We also 
observe that the $F_2$ and $G_P$ form factors make substantial 
contributions. Fig.~\ref{fig_pheno_rat} shows that the ratio of 
the axial-vector and vector correlation functions is nevertheless 
close to the value for a point-like nucleon, $g_A/g_V\simeq 1.25$.
We observe that the ratio of point-to-point correlation functions 
approaches this value from above. This is related to the fact 
that the axial radius of the nucleon is smaller than the vector 
radius. As a consequence, the point-to-point correlation function
at finite separation $\tau$ ``sees'' a larger fraction of the 
axial charge as compared to the vector charge.

\section{Instanton contribution to quark three-point functions}
\label{sec_app5}

 In this appendix we provide the results for the traces 
that appear in the single instanton contribution to the 
quark three-point function. Our starting point is the 
expression 
\be 
(P^{con}_{A/VQQ})^{\mu\nu} =
  P_{NZNZ}^{\mu\nu} + c_{A/V}
   \left(P^{\mu\nu}_{ZMm} + P^{\mu\nu}_{mZM}
   \right),
\ee
see equ.~(\ref{pi_3pt_z/nz}). Due to the Dirac structure of the non-zero 
mode part of the propagator, the $NZNZ$ term is the same for both 
the vector and axial-vector correlation functions. It has 4 parts 
stemming from combinations of the two terms of non-zero mode propagator 
equ.~\ref{NZ_prop}
\bea
P_{NZNZ_{11}}^{\mu\nu} &=& \frac{2H(x,z,y)(x-z)^\alpha(z-y)^\beta}
  {\pi^4(x-z)^4(z-y)^4}
S^{\nu\alpha\mu\beta}
\left[
  1+\frac{\rho^2}{z^2}
  \left(\frac{x\cdot z}{x^2} + \frac{z \cdot y}{y^2}
  \right) +
  \frac{\rho^4 x\cdot y}{x^2 z^2 y^2} 
\right] ,\\
P_{NZNZ_{12}}^{\mu\nu} &=& \frac{H(x,z,y)(x-z)^\alpha y^\beta}
   {2\pi^4(x-z)^4(z-y)^2z^2 y^2}
  [(z^{\alpha_0} + \frac{\rho^2}{x^2} x^{\alpha_0} ) 
  S^{\alpha_0\sigma\sigma_0\beta}
  \pm \frac{\rho^2}{x^2}x^{\alpha_0}\epsilon^{\alpha_0\sigma\sigma_0\beta}
  ]\nonumber\\
& &\times\left[
   \frac{\rho^2(z-y)^{\sigma_0}}{\rho^2+z^2}(S^{\nu\alpha\mu\sigma}
    \pm\epsilon^{\nu\alpha\mu\sigma})
    +\frac{\rho^2(z-y)^{\sigma}}{\rho^2+y^2}(S^{\nu\alpha\mu\sigma_0}
   \mp\epsilon^{\nu\alpha\mu\sigma_0})
\right],\\
P_{NZNZ_{21}}^{\mu\nu} &=& \frac{H(x,z,y)(z-y)^\alpha x^\beta}
   {2\pi^4(x-z)^2(z-y)^4z^2 x^2}
   [(z^{\alpha_0} + \frac{\rho^2}{y^2} y^{\alpha_0} ) 
   S^{\beta\sigma\sigma_0\alpha_0}
   \pm \frac{\rho^2}{y^2}y^{\alpha_0}\epsilon^{\beta\sigma\sigma_0\alpha_0} ]
 \nonumber \\
& &\times\left[
 \frac{\rho^2(x-z)^{\sigma_0}}{\rho^2+x^2}
  (S^{\nu\sigma\mu\alpha}\pm\epsilon^{\nu\sigma\mu\alpha})
  +\frac{\rho^2(x-z)^{\sigma}}{\rho^2+z^2}
   (S^{\nu\sigma_0\mu\alpha}\mp\epsilon^{\nu\sigma_0\mu\alpha})
    \right],\\
P_{NZNZ_{22}}^{\mu\nu} &=& 
    \frac{H(x,z,y) x^{\alpha_0} y^{\beta} }
  {4\pi^4(x-z)^2(z-y)^2 z^2 y^2 x^2}
  \frac{\rho^4}{(\rho^2+z^2)}
  \left[\frac{(x-z)^\sigma (z-y)^{\sigma_1}}{(\rho^2+x^2)}
(S^{\nu\alpha\mu\alpha_1} \pm \epsilon^{\nu\alpha\mu\alpha_1}) 
\right. \nonumber\\
& &+ \left.
\frac{(x-z)^\alpha (z-y)^{\alpha_1}}{(\rho^2+y^2)}
(S^{\nu\sigma\mu\sigma_1} \mp \epsilon^{\nu\sigma\mu\sigma_1}) 
\right]
 T_{\pm}[\alpha_0,\alpha,\sigma,\alpha_1,\sigma_1,\beta],
\eea
where 
\be
H(x,z,y)=\left\{  \left( 1+\frac{\rho^2}{z^2} \right)
 \sqrt{ \left(1+\frac{\rho^2}{x^2}\right) 
        \left(1+\frac{\rho^2}{y^2}\right) }  \right\}^{-1}
\ee
and the Dirac trace $T_{\pm}$ is defined as
\bea
\label{DTrace6mp}
 T_{\mp}[\mu,2,3,4,5,6] &\equiv&
 \left[\left(
  g^{\mu 2}S^{3456} - g^{\mu 3}S^{2456} + g^{\mu 4} S^{2356} - g^{\mu 5
  }S^{2346} + g^{\mu 6} S^{2345}
 \right)
\right. \nonumber \\
&& \hspace{-2.2cm}\mp \left.  \left(
  g^{\mu 2}\epsilon^{3456} - g^{\mu 3}\epsilon^{2456} +
  g^{23}\epsilon^{\mu 456} + g^{45}\epsilon^{\mu 236} -
  g^{46}\epsilon^{\mu 235} + g^{56}\epsilon^{\mu 234}
 \right)
\right],
\eea
where $2,3,\ldots$ is short for $\mu_2,\mu_3,\ldots$.
The $S_{ZM}S_m$ term is easily seen to be
\bea
P^{\mu\nu}_{ZMm}&=&{\rm Tr}[-\Psi_0(x)\Psi_0^{+}(z)\gamma^{\mu}
(-\Delta_{\pm}(z,y)\gamma_{\pm})\gamma^{\nu}]\nonumber\\
&=& \frac{\varphi(x)\varphi(z)x^{\alpha_0}z^{\beta_0}}
{8\pi^2(z-y)^2}
\frac{1}
{\sqrt{(1+\frac{\rho^2}{z^2})(1+\frac{\rho^2}{y^2})}}
T_{\mp}[\alpha_0,\sigma_1,\sigma,\beta_0,\mu,\nu]
\nonumber\\
& & \times
\left[
\delta^{\sigma_1,\sigma}+\frac{\rho^2z^\alpha y^\beta}{z^2y^2}
(S^{\sigma_1\sigma\alpha\beta}\pm\epsilon^{\sigma_1\sigma\alpha\beta}
 )\right]
\eea
with $\varphi(x)=\rho/(\pi\sqrt{x}(x^2+\rho^2)^{3/2})$.
The $S_mS_{ZM}$ term is obtained similarly
\bea
P^{\mu\nu}_{mZM}&=&
{\rm Tr}[(-\Delta_{\pm}(x,z)\gamma_{\pm})\gamma^{\mu}
(-\Psi_0(z)\Psi_0^{+}(y))\gamma^{\nu}]\nonumber\\
&=& \frac{\varphi(z)\varphi(y)z^{\alpha_0}y^{\beta_0}}
{8\pi^2(x-z)^2}
\frac{1}
{\sqrt{(1+\frac{\rho^2}{x^2})(1+\frac{\rho^2}{z^2})}}
T_{\mp}[\alpha_0,\sigma_1,\sigma,\beta_0,\nu,\mu]
\times
\nonumber\\
& & \times
\left[
\delta^{\sigma_1,\sigma}+\frac{\rho^2x^\alpha z^\beta}{x^2z^2}
(S^{\sigma_1\sigma\alpha\beta}\pm\epsilon^{\sigma_1\sigma\alpha\beta}
 )\right].
\eea
For all the above formulas, we need to add instanton and anti-instanton
contribution and integrate over the position of the instanton, which was 
suppressed above. As usual, for an instanton at position $z_I$ we have
to shift $x,y,z$ according to $x\to (x-z_I)$, etc.

\newpage\noindent

\end{document}